\documentclass[final,12pt, authoryear]{elsarticle}

\usepackage{url}
\usepackage[margin=1.1in]{geometry}
\usepackage{float}
\usepackage[hidelinks,colorlinks=true,linkcolor=blue,citecolor=blue]{hyperref}
\usepackage{amsmath}
\usepackage[graphicx]{realboxes}
\usepackage{adjustbox}
\usepackage[table,xcdraw]{xcolor}
\usepackage[T1]{fontenc}
\usepackage{epsf}
\usepackage{caption}
\usepackage{subcaption}
\newcommand{\probP}{\text{I\kern-0.15em P}}
\usepackage{pifont}
\usepackage{newunicodechar}
\newunicodechar{☑}{\ding{51}}
\newunicodechar{✗}{\ding{55}}

\usepackage{amssymb}

\journal{}

\begin{document}

\begin{frontmatter}



\title{A Novel Approach to Queue-Reactive Models: The Importance of Order Sizes}



\author[inst1,inst2]{Hamza Bodor}
\ead{bodor.hamza@gmail.com}

\author[inst2]{Laurent Carlier}
\ead{laurent.carlier@bnpparibas.com}

\affiliation[inst1]{organization={Université Paris 1 Panthéon-Sorbonne, Centre d’Economie de la Sorbonne},
            addressline={106 Boulevard de l'Hôpital}, 
            city={Paris Cedex 13},
            postcode={75642},
            country={France}}

\affiliation[inst2]{organization={BNP Paribas Corporate and Institutional Banking, Global Markets Data \& Artificial Intelligence Lab},
            addressline={20 boulevard des Italiens}, 
            city={Paris},
            postcode={75009},
            country={France}}

\begin{abstract}
In this article, we delve into the applications and extensions of the queue-reactive model for the simulation of limit order books. Our approach emphasizes the importance of order sizes, in conjunction with their type and arrival rate, by integrating the current state of the order book to determine, not only the intensity of order arrivals and their type, but also their sizes. These extensions generate simulated markets that are in line with numerous stylized facts of the market. Our empirical calibration, using futures on German bonds, reveals that the extended queue-reactive model significantly improves the description of order flow properties and the shape of queue distributions. Moreover, our findings demonstrate that the extended model produces simulated markets with a volatility comparable to historical real data, utilizing only endogenous information from the limit order book. This research underscores the potential of the queue-reactive model and its extensions in accurately simulating market dynamics and providing valuable insights into the complex nature of limit order book modeling.
\end{abstract}



\begin{keyword}
Limit order book \sep Market simulators \sep Stylized facts \sep Market microstructure
\end{keyword}

\end{frontmatter}


\newpage

\section{Introduction}

Limit order book (LOB) simulation is a critical area of study with far-reaching implications in the field of Finance. It serves as the backbone for understanding and predicting market dynamics, thereby aiding in the development of effective trading strategies \mbox{\citep{Cartea2015}}. The primary utility of LOB simulation is manifested in its application to the development and assessment of algorithmic trading strategies. By simulating various market scenarios and order book dynamics, LOB models provide crucial insights that enable traders to formulate strategies that are responsive to different market conditions. This not only aids in understanding emerging market trends but also plays a significant role in enhancing the decision-making processes in algorithmic trading \mbox{\citep{cohen2012limit, balch2019evaluate}}. Moreover, it plays a pivotal role in risk management, where understanding the LOB can help mitigate potential losses due to market volatility \citep{ABIDES}. 
LOB simulation aims to accurately replicate the complex dynamics of financial markets at the microscopic scale. This involves capturing the intricate interactions between different market participants, the impact of external events on market behavior, and the inherent uncertainty and randomness in market movements. However, achieving a high degree of accuracy in LOB simulation is a complex task. The intrinsic complexity of financial markets, coupled with the vast amount of data involved and the need for high computational efficiency, makes this a challenging problem. 

 In the realm of limit order book simulation, a variety of methods have been developed, each offering distinct advantages and challenges. Traditional approaches often rely on theoretical models like Markovian Poisson processes \citep{avellaneda2008high, cont2010stochastic, huang2015simulating} and Hawkes processes \citep{toke2011, lu2018high, fosset2022non, wu2022queue}. These methods are primarily used to model the pattern and impact of order arrivals in LOBs. Rooted in the concept of regular patterns in order flows, these models leverage statistical assumptions to yield solutions that are concise and analytically tractable.

Alternatively, agent-based methods model the limit order book as a system of interacting agents, each following a set of rules or strategies. These methods can capture the heterogeneity of market participants and the strategic interactions between them. Abides \citep{ABIDES}, an interactive agent-based discrete event simulation environment, has encouraged multiple works on agent-based limit order book generation \citep{ vyetrenko2020get, amrouni2021abides}. More recently, \citet{lussange2021modelling} propose an agent composed of two parts: one part predicts the asset price based on the state of the order book, and the second component uses this prediction to decide on the action to take. They demonstrate that a configuration with agents of this type, along with Zero Intelligence (ZI) agents, allows for the production of a simulated market faithful to multiple stylized facts of the equity market.

Recently, machine learning techniques, particularly Generative Adversarial Networks (GANs), have also been applied to limit order book simulation \citep{sirignano2019universal, zhang2019integrating, coletta2023conditional}. These methods can learn complex patterns from data and generate realistic order book dynamics. They have shown promising results and good capability of taking into account different aspects of the limit order book. 

\section{Methodology}

\subsection{Queue-Reactive Model}

The Queue-Reactive (QR) model of \citet{huang2015simulating} has gained significant attention from numerous researchers due to its simple yet effective simulation framework for the limit order book. This model verifies multiple stylized facts and addresses various issues that other basic simulators may encounter \citep{mariotti2023zero}.

In \citet{huang2015simulating}, the limit order book is viewed as a $2K$-dimensional vector, where $K$ represents the depth of the LOB. The reference price, often\footnote{The reference price typically matches the mid-price when the spread is an even multiple of the tick size. Otherwise, it is adjusted to either half a tick above or below the mid-price, depending on the specific market conditions. For more details on this mechanism and its implications on market dynamics, see subsequent paragraphs.} the mid-price, divides the LOB into two halves: the bid side and the ask side. Each side is populated with an aggregation of limit orders, represented by $Q_{\pm i}$, where $i$ denotes the distance from the reference price in terms of ticks, the smallest price increment change an asset can make on the market.

The quantities at each limit are subject to changes due to three types of orders: \emph{Limit}~(L) orders, which increase the size of the queue; \emph{Cancel} (C) orders, which remove an existing order; and \emph{Market} (M) orders, which immediately match with an existing limit quantity, resulting in trades (hence, this type may also be referred to as \emph{Trade} orders).  

In the queue-reactive model, the arrival of order flow at a given price level is modeled as a heterogeneous Poisson process. The intensity of this process depends solely on the current state of the order book, particularly the queue sizes. The intensities of limit, cancel, and market order arrivals are denoted by $\lambda^L$, $\lambda^C$, and $\lambda^M$ respectively, and these intensities are functions of the queue size at each limit.

\begin{figure}[H]
\centering
\includegraphics[width = 0.55\textwidth]{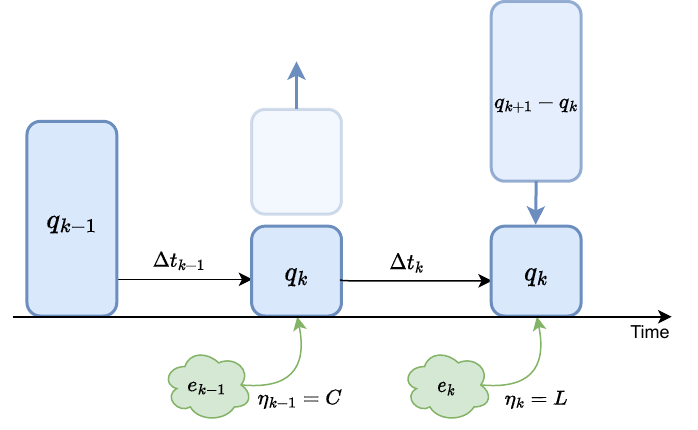}
\caption{Illustration of queue size evolution. This diagram represents the sequential updates in the size of a given queue over time. Initially, the queue size is $q_{k-1}$. It is then updated: following a time interval of $\Delta t_{k-1}$, the event $e_{k-1}$, a cancellation, reduces the queue size. Subsequently, after a time interval $\Delta t_k$, a new event, $e_k$ (characterized as a limit order), occurs, injecting additional liquidity and increasing the queue size by $q_{k+1} - q_k$. These events are identified by their corresponding order types: $\eta_{k-1} = C$ for cancellation and $\eta_k = L$ for a limit order, affecting the queue's liquidity state.}
\label{fig:QR_schema}
\end{figure}

Figure \ref{fig:QR_schema} shows how the queue size of a given level is updated. where event $k-1$ decreases the size of the queue from $q_{k-1}$ to $q_k$ after $\Delta t_{k-1}$. Similarly, the size of the queue increases to $q_{k+1}$ after $\Delta t_{k}$ when event $k$ occurs.

\begin{figure}[H]
\centering
\includegraphics[width = \textwidth]{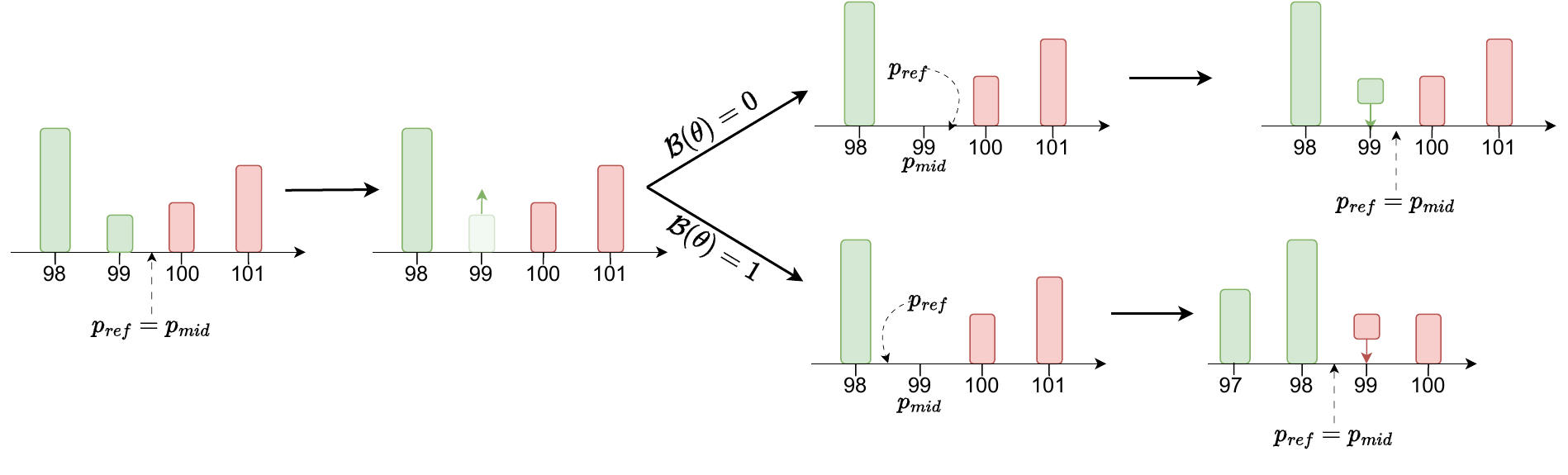}
\caption{Transition of the reference price \( p_{ref} \) in response to spread changes. When the spread is an even multiple of the tick size, \( p_{ref} \) can adjust by \(p_{mid} \pm 0.5\) tick. The figure illustrates this behavior, showing that with probability \( \theta \), the reference price shifts, enabling order fulfillment at the new price level. \( \mathcal{B}(\theta) \) indicates a Bernoulli trial with success probability \( \theta \), determining the reference price change scenario.}
\label{fig:QR_price_change}
\end{figure}

While this explains how the queue sizes evolve under the queue-reactive model, it is also important to understand how the price changes under this model. In fact, Figure \ref{fig:QR_price_change} displays a scenario of price change under the QR model where this occurs when the best bid or the best ask price is totally consumed (either after a cancel or a market order). When this happens and the spread value is an even multiplier of the tick size, the reference price $p_{ref}$ will now have two possible values: $p_{mid} + 0.5~\textrm{tick}$ or $p_{mid} - 0.5~\textrm{tick}$. The model postulates then that with probability $\theta$, the reference price changes value, allowing the opposite side to fill the new price. Otherwise, the reference price stays at its value, and the levels do not change. The parameter $\theta$ is calibrated in order to reproduce the mean reversion ratio, as discussed in \citet{huang2015simulating}.

For a set $\mathcal{E}$ of events at a given queue $i$, $\mathcal{E} = (e_k = (\eta_k, q_k, \Delta t_k))_k$, the likelihood of the QR model can be expressed as:

\begin{equation}
\mathcal{L}( \mathbf{\lambda} | \mathcal{E} ) = \prod_{k} e^{-\lambda^{\text{global}}(q_k)\Delta t_k}\lambda^{\eta_k}(q_k)
\label{eq:likelihood}
\end{equation}

where:

\begin{itemize}
\item $\eta_k$ is the type of the $k^{th}$ event, $\eta \in \mathcal{T}, \mathcal{T} = \{L, C, M\} $ is the set of possible event types.
\item $q_k$ is the size of the queue just before the $k^{th}$ event.
\item $\Delta t_k$ is the time difference between events $k-1$ and $k$.
\item $\lambda^{\text{global}}(q_k) = \sum_{\eta \in \mathcal{T}} \lambda^\eta(q_k)$, the intensity of arrival of an order when the size of the queue is $q_k$.
\end{itemize}

\newpage
For a given size $n$ of the queue, measured in Average Event Size\footnote{This is the average of all events sizes occurring regardless of their type. \cite{huang2015simulating} propose also the usage of Average Trade Size (ATS) as a normalizing factor, which accounts only for the size of trade events.} (AES), maximum likelihood estimators are known in closed form:

\begin{align*}
\hat{\Lambda}(n) &= \left( \frac{1}{\#\{k \mid q_k = n\}} \sum_{\{k \mid q_k = n\}} \Delta t_k \right)^{-1} \\
\hat{\lambda}^L(n) &= \hat{\Lambda}(n) \frac{\# \{e_k \in \mathcal{E} ~|~ \eta_k = L, q_k = n\}}{\# \{e_k \in \mathcal{E} ~|~ q_k = n\}} \nonumber\\
\hat{\lambda}^C(n) &= \hat{\Lambda}(n) \frac{\# \{e_k \in \mathcal{E} ~|~ \eta_k = C, q_k = n\}}{\# \{e_k \in \mathcal{E} ~|~ q_k = n\}} \nonumber\\
\hat{\lambda}^M(n) &= \hat{\Lambda}(n) \frac{\# \{e_k \in \mathcal{E} ~|~ \eta_k = M, q_k = n\}}{\# \{e_k \in \mathcal{E} ~|~ q_k = n\}} \nonumber \\
\end{align*}

In our study, we base our experimental results on Bund futures data. Bund futures are well-known for their high liquidity and the abundance of available data, making them ideal for detailed analysis. The Bund is a large tick asset, for which the spread is usually of one tick size, which helps reduce the impact of large spreads on model behavior. This is quite different from the assets used in \cite{huang2015simulating}, allowing us to test the model in a market setting with different characteristics.

\begin{figure}[H]
\centering
\includegraphics[width = 0.8\textwidth]{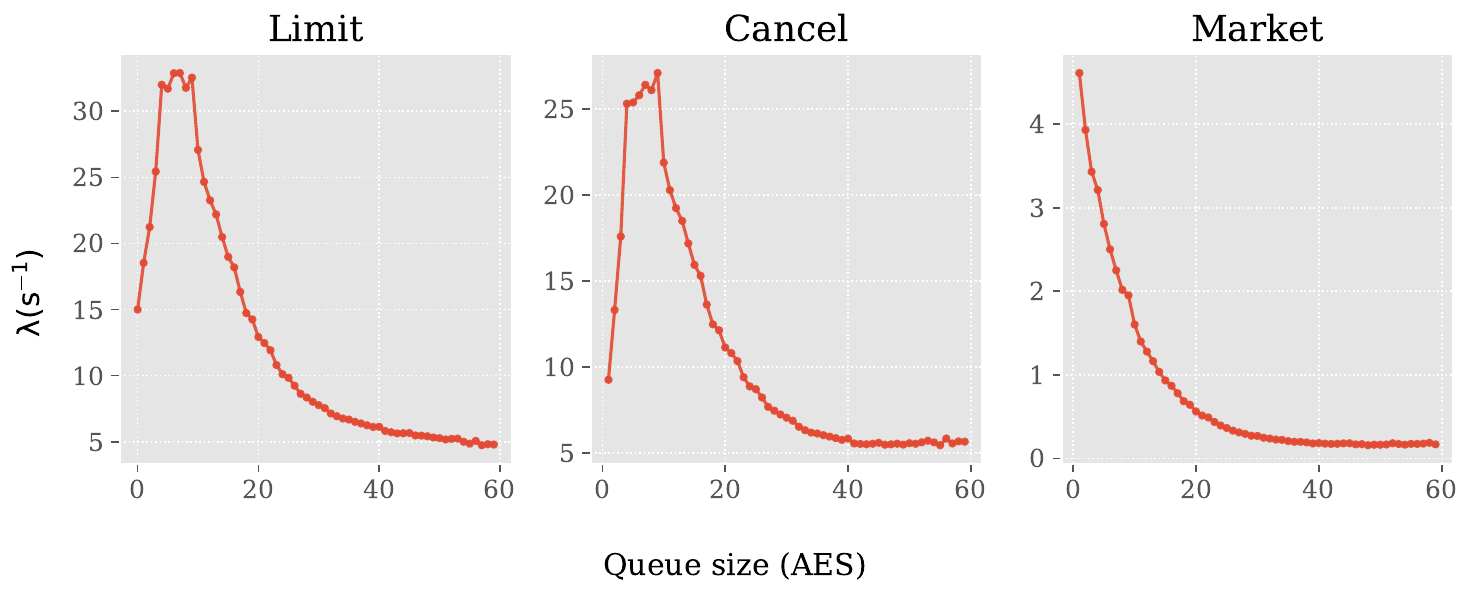}
\caption{Results of queue-reactive model on Bund futures - Intensities as functions of queue size (best levels).}
\label{fig:QR_vanilla_intensities}
\end{figure}

Figure \ref{fig:QR_vanilla_intensities} shows the calibrated intensities on our data, one can conclude that:

\begin{itemize}
\item Intensities vary significantly with the size of the queue, which justifies the choice of this model over a simple Poisson model.
\item The limit intensity generally decreases, except when queue sizes are \emph{tiny}. This phenomenon can be understood by considering the behavior of Market Participants~(MPs). MPs may perceive it as risky to enter a level with an extremely low volume, until a certain threshold (typically between 7 AES to 10 AES) is reached. At this threshold, there is a noticeable spike in limit orders, which can be attributed to the MPs' pursuit of priority. However, as the size of the queue grows, this pursuit diminishes.

\item The same behavior is observed for cancel orders. However, it has the opposite interpretation; this may be because MPs who do not want to be executed withdraw their orders when the size of the queue is around 10 AES. After that, market participants generally prefer to be executed rather than cancel and move to a deeper level. This contradiction in behavior with the limit intensity may indicate that there are two major types of market participants: one type always seeks to be executed as quickly as possible without paying the spread (without being aggressive), and others who prefer to seek priority without being executed, possibly waiting for better opportunities. This hypothesis of a fragmented market has also been adopted by other research works such as \citet{de1995exchange} or \citet{lux1998scaling}.

\item The trading rate decays with the available volume at the best level. This phenomenon is easily explained by market participants \emph{rushing for liquidity} when it is scarce, and \emph{waiting for a better price} when liquidity is abundant, similar results are noticed on the equity markets in the \citet{huang2015simulating} study.
\end{itemize}

\begin{figure}[H]
\centering
\includegraphics[width = 0.8\textwidth]{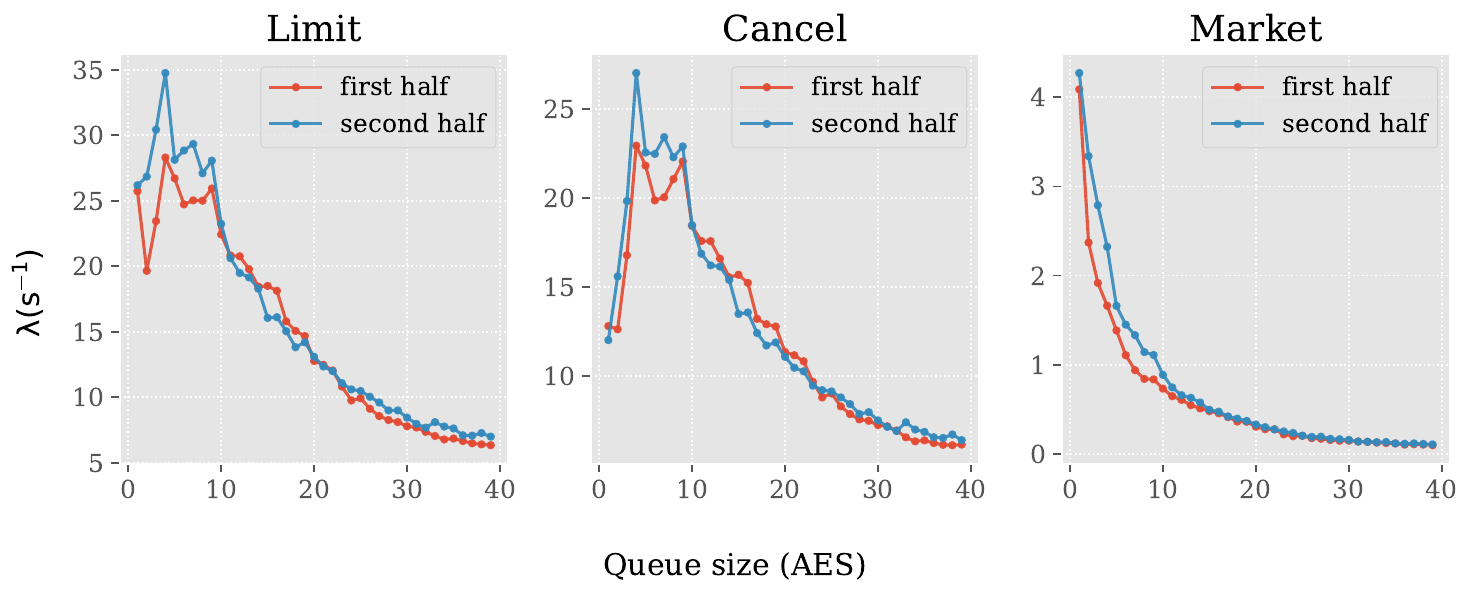}
\caption{Intensities of limit, cancel, and market orders after the dataset is divided into two equal time segments.}
\label{fig:QR_vanilla_two_halves}
\end{figure}

Moreover, the dataset, encompassing an entire year's worth of  data, was evenly split into two halves, ensuring the chronological sequence of trading days was preserved. This deliberate division was undertaken to examine the stability and reliability of the model calibrations over two distinct time periods. After calibrating the models for each interval separately, we analyzed the resulting order flow intensities for limit, cancel, and market orders relative to queue sizes. The outcomes, illustrated in Figure \ref{fig:QR_vanilla_two_halves}, reveal a noteworthy consistency across both periods, thereby validating the temporal stability of the calibrated intensities within the span of the examined period.

Now that we have the intensity of arrival of orders, and their types, the last and only missing information about the order is its size. We explore two methods of orders sizing:

\begin{itemize}
\item Each order is assigned a uniform size, determined by rounding up the average event size ($\lceil AES_i \rceil$) for each level $i$. This approach, referred to as \textbf{QRU} (QR model with Unitary order sizes), aligns with the model used in \citep{huang2015simulating}.
\item The size of each order is sampled from the stationary distribution of order sizes for its respective level, denoted as \textbf{QR}.
\end{itemize}

\subsection{Queue-Reactive Model Extensions}

In addition to these two variants of the QR model, we propose two other extensions heavily inspired by it. The goal is to examine the impact of incorporating a more sophisticated model for order sizes on the simulator performances. The first extension involves a modified QR model that includes five possible event types, by adding two additional types, \textit{cancel\_all} and \textit{market\_all}, that model the complete consumption of the queue. This aims to simulate instances where the size of a consumption order precisely matches the queue size. The other extension involves incorporating the order size as a variable in determining the nature of order arrival, similar to its influence by event type.


\subsubsection{Five-Type Queue-Reactive model (\textbf{FTQR})}

As mentioned earlier, and as will be discussed in section \ref{sec:results}, one of the main limitations of the queue-reactive model is its inability to reproduce historical market volatility. This is primarily due to a slower progression of the LOB when the size of one of the best queues is small, compared to historical progression. In fact, as the sizes of the orders are drawn from the stationary distribution of sizes, which is very skewed towards small sizes, it leads to a very slow decrease of the queue, hence, fewer price changes and low volatility. 



\begin{figure}[H]
\centering
\begin{subfigure}{0.45\textwidth}
    \includegraphics[width = 1\textwidth]{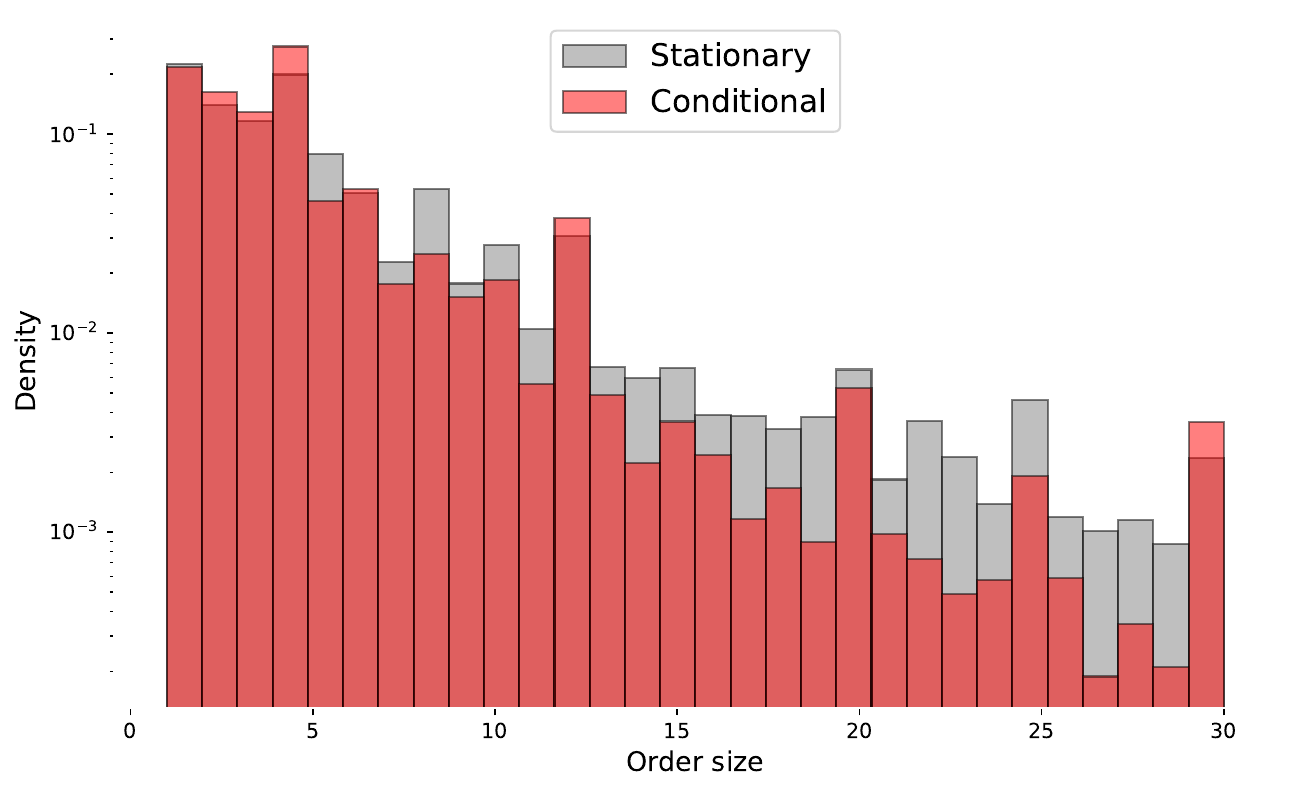}
    \caption{Cancel orders.}
    \label{fig:first3}
\end{subfigure}
\hfill
\begin{subfigure}{0.45\textwidth}
    \includegraphics[width = \textwidth]{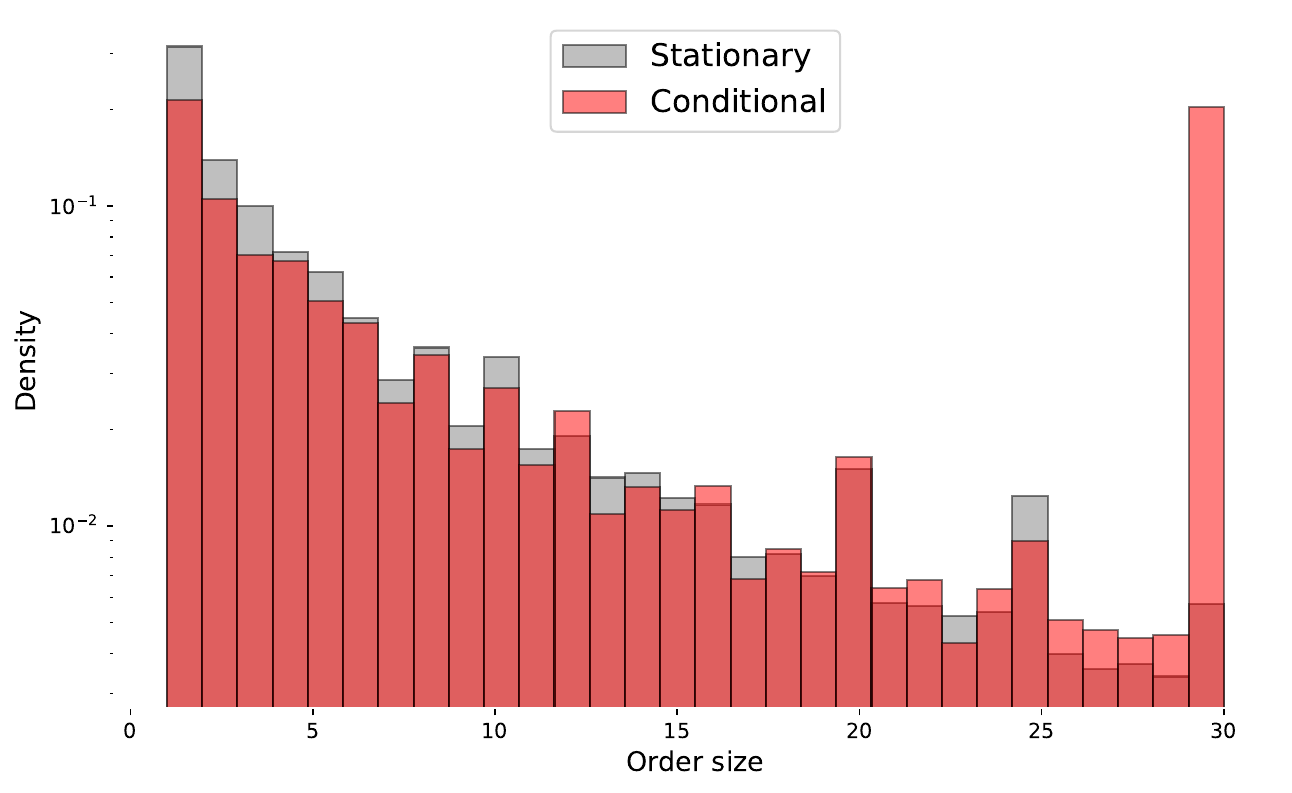}
    \caption{Market orders.}
    \label{fig:second3}
\end{subfigure}
\hfill

\caption{Comparison of stationary distribution of order sizes up to 30 (in gray) with conditional distribution when queue size equals 30 (in red), presented on a logarithmic scale.}

\label{fig:hist_percentage_order_size}
\end{figure}

In fact, Figure \ref{fig:hist_percentage_order_size} presents the stationary distribution of order sizes truncated at 30 lots \footnote{In the context of BUND Futures, a queue comprises multiple contracts (lots), each valued at 100,000 Euros.}($\approx 5 AES$) (\textit{stationary}), and the conditional distribution of order sizes when the size of the queue is 30 lots (\textit{conditional}). It is an example of the state of order sizes when the queue size is relatively low. This illustration highlights the disparity between the two distributions, with the conditional distribution leaning towards larger values and, correspondingly, greater queue consumption.

In particular, there is a pronounced surge of orders that deplete 100\% of the queue, predominantly for market orders. The frequency of such orders is more than 40 times higher than that in the stationary queue size consumption ratio. This significant discrepancy provides an explanation as to why queue consumption is markedly slower when stationary sizes are utilized. Especially after showing the link between queue consumption and price dynamics in Figure \ref{fig:QR_price_change}. This, in turn, elucidates why the simulated price exhibits considerably less volatility compared to actual prices.

These insights motivate a first extension of the QR model that take into account the phenomenon of complete queue consumption when the size of the queue is small. In such situations, two specific events tend to occur:


\begin{itemize}
\item Market participants seize the available quantity and aggressively execute it (using market orders).
\item Market participants who possess these scarce quantities feel uncomfortable being alone in the queue and cancel all their orders. Another way of seeing it is that when market participants are alone in the queue, it is \emph{free} to cancel their order without worrying about losing priority in the queue.
\end{itemize}

In this context, we propose adding two additional events that correspond to the complete consumption of available quantity, either through trades or cancellations. The likelihood of this new model is similar to that presented in Equation \ref{eq:likelihood} by including \emph{market\_all} and \emph{cancel\_all} events in $\mathcal{T}$, the set of possible event types.

\begin{figure}[H]
\centering
\includegraphics[width = 0.8\textwidth]{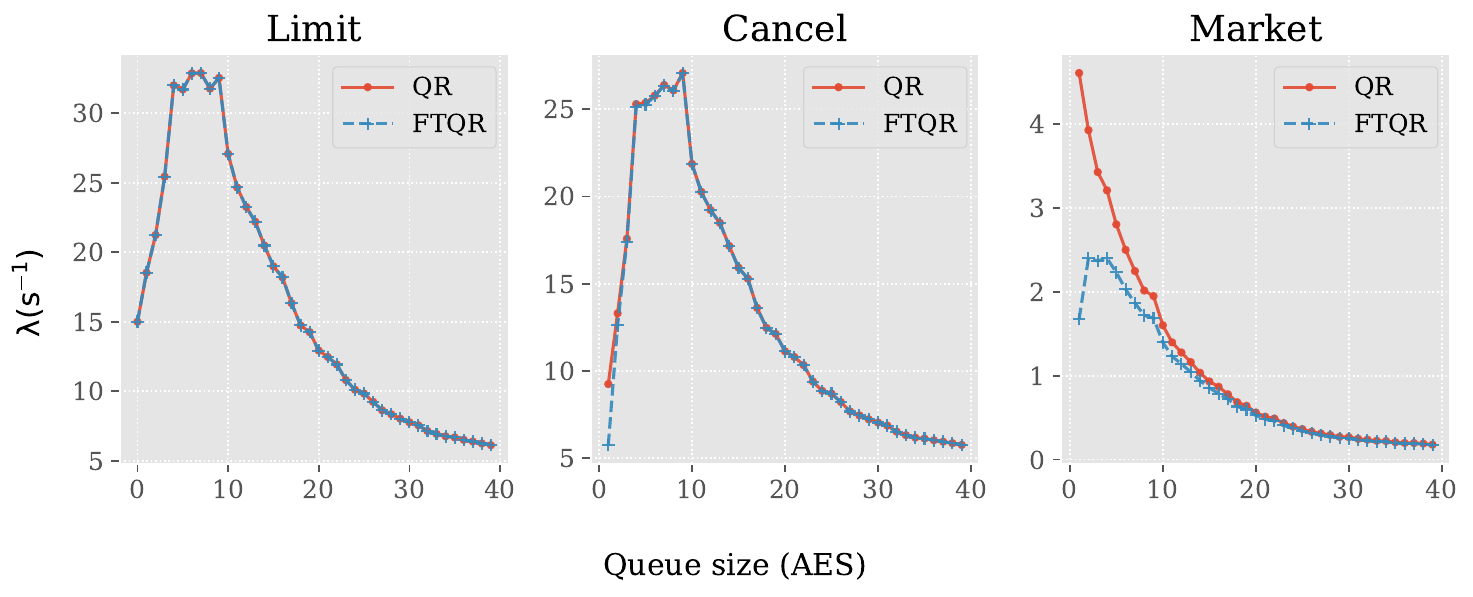}
\caption{Comparison of QR and FTQR intensities on Bund future data.}
\label{fig:FTQR_QR}
\end{figure}

\begin{figure}[H]
\centering
\includegraphics[width = 0.5\textwidth]{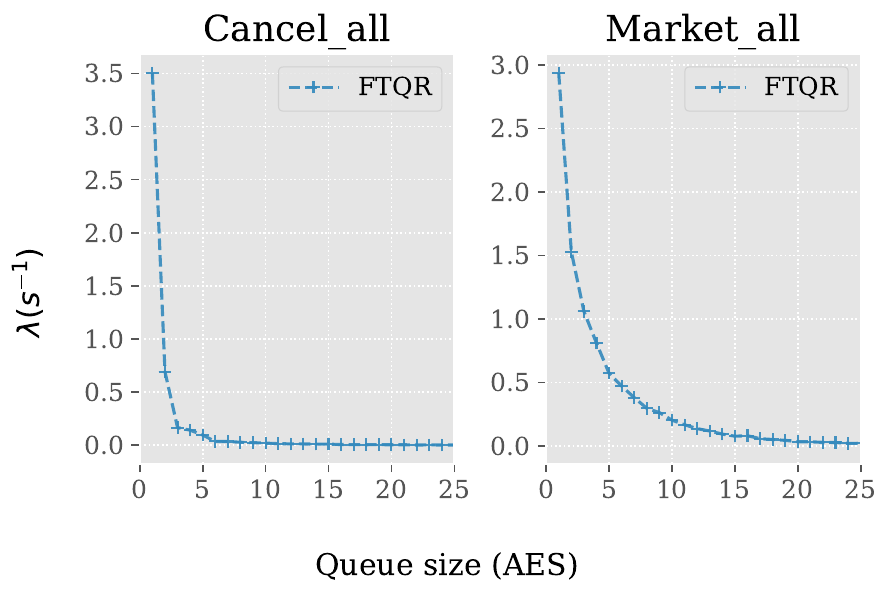}
\caption{Calibrated intensity functions for additional types of the FTQR model applied to Bund future data, truncated at values less than 25 lots. Intensities for larger queue sizes are negligible.}

\label{fig:additional_types}
\end{figure}

Figures \ref{fig:FTQR_QR} and \ref{fig:additional_types} above show the obtained intensities of the five-types queue-reactive model and compare them with the vanilla QR model intensities, revealing the following observations:

\begin{itemize}
\item \textbf{Limit orders:} The limit order intensities are identical in both models, as the adjustment does not affect limit orders.
\item \textbf{Cancel orders:} The cancel order intensities are also the same, except for smaller queue sizes where the intensities of the adjusted model are lower. This difference originates from the additional cancel type that consumes the entire level. Nevertheless, the impact of such cancel orders on overall liquidity is negligible compared to other cancel orders.
\item \textbf{Market orders:} This order type is the most influenced by the adjustment. We notice that the new intensities are lower than the original ones, particularly for smaller queue sizes, as these situations provide more opportunities for rarity chasing.
\item \textbf{Market\_all orders:} The new intensities compensate for the earlier ones, particularly for small queue values, reflecting a significant value. This finding indicates that many orders consume liquidity at a high rate when this liquidity is scarce. The intensity of market\_all orders decays with the queue size.
\item \textbf{Cancel\_all orders:} This order type exhibits a behavior similar to the previous one. It is mainly interpreted as the intensity of arrival of events where a market participant, who is alone in the queue, completely withdraws from the market.
\end{itemize}

\subsubsection{Size-Aware Queue-Reactive model (\textbf{SAQR})}


In this section, we explore the impact of order sizes on the dynamics of limit order book simulations. While research has extensively analyzed the rates and types of event arrivals, the size of orders has received less attention.

To build upon this dimension, we propose an adaptation of the QR model that takes into account the size of orders. We achieve this by changing the possible values of $\mathcal{T}$ in Equation (\ref{eq:likelihood}) to a 2-dimensional order type, $\mathcal{T} = \{L, C, T\} \times \mathbb{N}^*$. This means that the intensity nature not only specifies the type of the order but also its size. This adjustment does not change the formulation of the likelihood. And the solution, for $(\eta, s)$ and a queue size $n$, is:

$$ \hat{\lambda}^{(\eta, s)}(n) = \hat{\Lambda}(n) \frac{\# \{e_k \in \mathcal{E} ~|~ \eta_k = \eta, s_k = s, q_k = n\}}{\# \{e_k \in \mathcal{E} ~|~ q_k = n\}} \nonumber $$
where $s_k$ is an additional model variable and represents the size of the $k^{th}$ order.

\begin{figure}[htbp]
\centering
\begin{subfigure}{0.6\textwidth}
    \includegraphics[width = 1\textwidth]{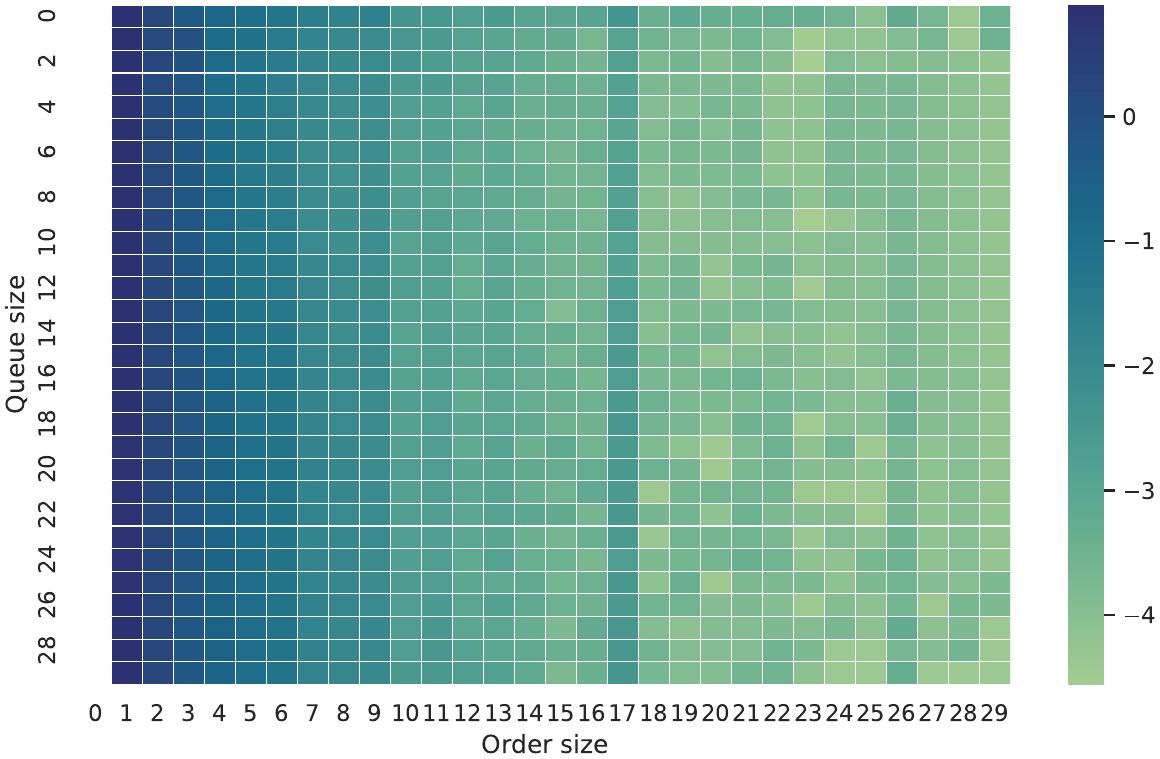}
    \caption{Limit orders}
    \label{fig:first1}
\end{subfigure}
\hfill
\begin{subfigure}{0.6\textwidth}
    \includegraphics[width = \textwidth]{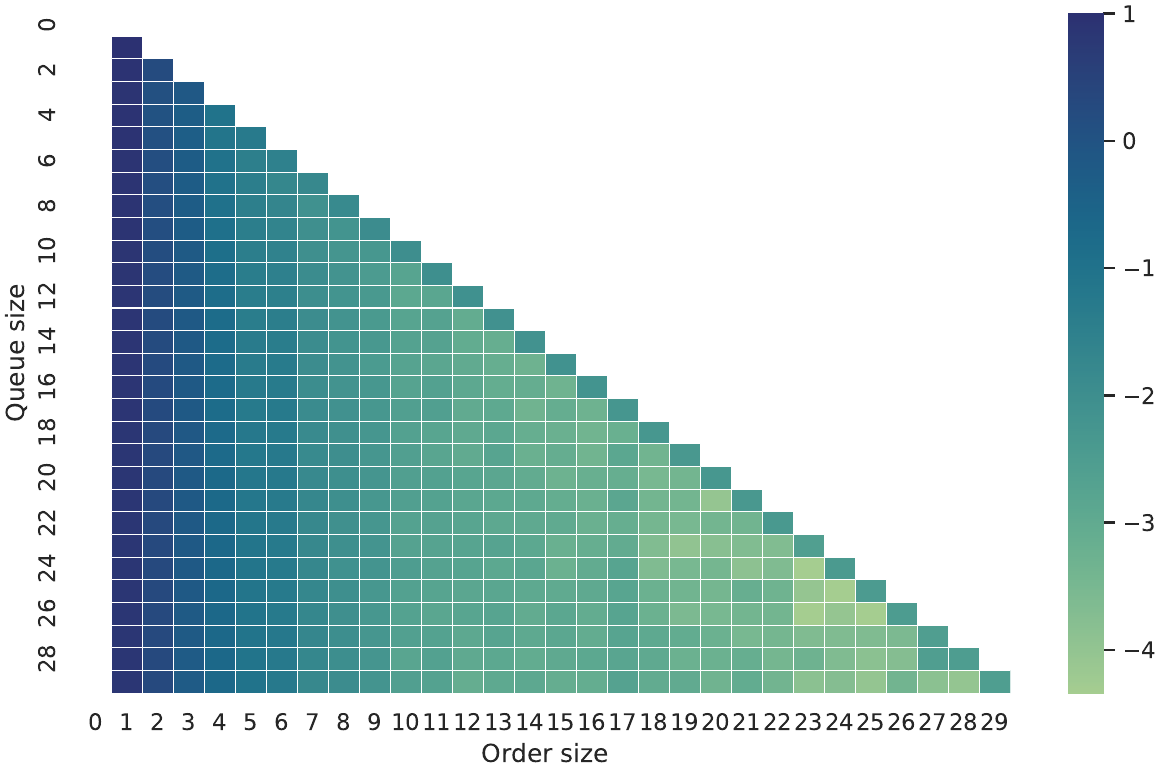}
    \caption{Cancel orders}
    \label{fig:second1}
\end{subfigure}
\hfill

\begin{subfigure}{0.6\textwidth}
    \includegraphics[width = \textwidth]{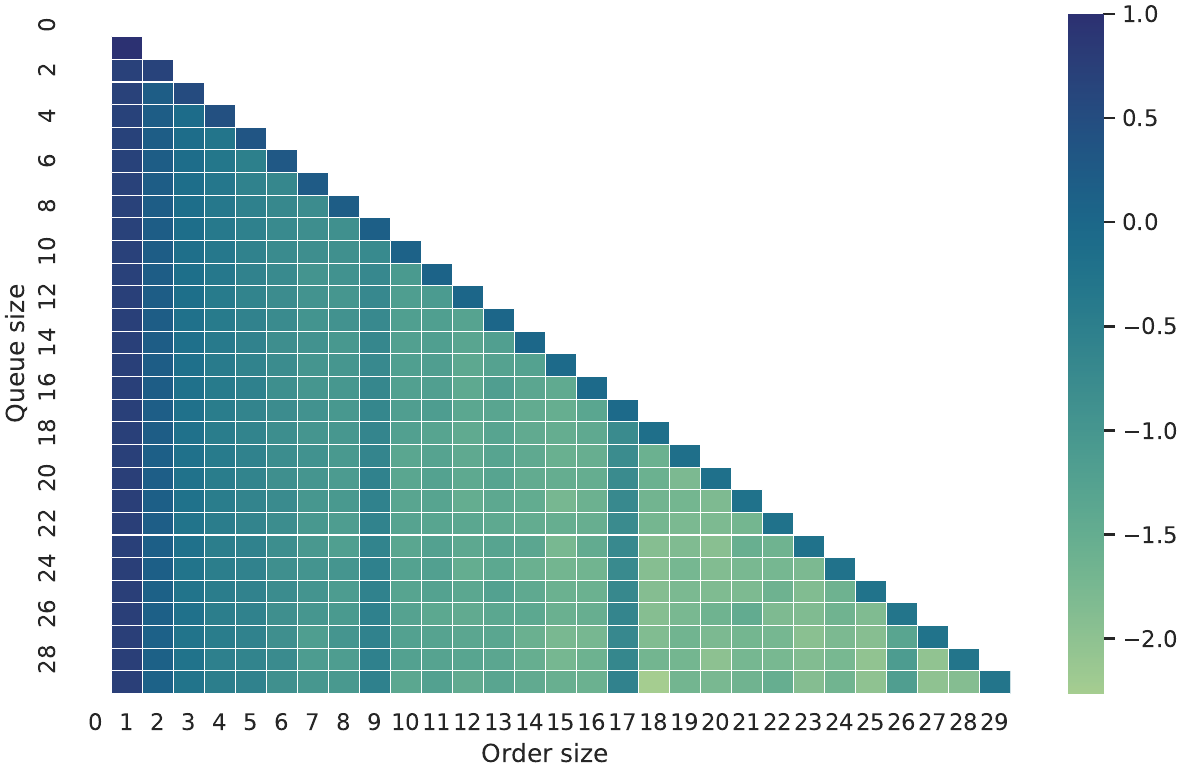}
    \caption{Market orders}
    \label{fig:second2}
\end{subfigure}
\hfill
        
\caption{Matrix heatmap of normalized calibrated intensities On Bund future data  (logarithmic scale). Normalization is done such that intensities sum to one for each row.}
\label{fig:Matrix_heatmap_of_normalized_intensities}
\end{figure}

Figure \ref{fig:Matrix_heatmap_of_normalized_intensities} presents the intensity matrix for the three types of orders: limit, cancel and market orders (averaged for each queue size). It is evident from the figure that while the distribution of limit order sizes remains relatively consistent across the different queue size, the same cannot be said for cancel and market orders. 

In particular, the diagonal values in the intensity matrix, which represent orders that consume the entire available liquidity, are noticeably dominant. This dominance is especially pronounced for market orders, confirming the tendency of these orders to consume the entire quantity available at the best levels.


Interestingly, this phenomenon is more prevalent when the available quantity is relatively low. This observation highlights the fact that the distribution of order sizes is not uniform, but rather varies significantly depending on the size of the queue. This nuanced understanding of order size distribution is crucial for accurately simulating and analyzing the dynamics of the limit order book. 

Furthermore, one can notice in the conditional market intensities some particularly \emph{salient} intensities at 9 AES, 17 AES and 26 AES. These are buckets including orders of sizes 50 lots, 100 lots and 150 lots, respectively. These round numbers are generally more occurrent in historical data than other order sizes, hence the high intensity of arrival of orders of these sizes.

\subsection{Hawkes Models}

Hawkes processes, named after Alan Hawkes \citep{hawkes1971spectra}, are a class of self-exciting and mutually-exciting point processes. Unlike Poisson processes, they dispense with the independence-of-increments property, allowing past events to influence the likelihood of future events. This characteristic makes them particularly adapted for modeling clustering and contagion phenomena, which are prevalent in social media networks, seismology, and financial markets.

Mathematically, a Hawkes process is defined by its intensity function, $\lambda(t)$, which depends on the entire history of the process. The intensity function is often formulated as:

\[ \lambda(t) = \mu + \int_{-\infty}^{t} h(t - s) dN(s) \]

where $\mu$ is the baseline intensity, $h$ is a memory kernel describing the influence of past events, and $N(s)$ is the count of events up to time $s$. 

The complexity of financial markets, with multiple interacting variables, necessitates a more comprehensive approach to capture these dynamics accurately. This leads us to the multivariate Hawkes process, which extends the basic concept to encompass multiple types of interdependent events.

Mathematically, a multivariate Hawkes process is defined by a set of intensity functions, $\lambda_i$, each depending on the entire history of the process. The intensity function for the \(i^{th}\) component of the process at time \(t\) can be expressed as:
\[ \lambda_i(t) = \mu_i + \sum_{j=1}^{n} \int_{-\infty}^{t} h_{ij}(t - s) dN_j(s) \]
where \(\mu_i\) is the baseline intensity for the \(i\)-th component, \(h_{ij}\) is a memory kernel describing the influence of past events of type \(j\) on the current rate of type \(i\), and \(N_j(s)\) is the count of type \(j\) events up to time \(s\). This model captures not only the self-excitations ($j = i$) of events, but also the cross-excitation ($j \ne i$) between events.

In the context of limit order book (LOB) modeling, Hawkes processes have been instrumental in describing the flow of orders. These models view different order types as distinct events in a multivariate point process, capturing the dynamics of market reactions to these orders and their interrelations. Significant developments in LOB modeling include contributions from \citet{zhao2010} and \citet{toke2011}, who have expanded upon the \mbox{\citet{cont2010stochastic}} framework. Zhao's empirical analysis of crude oil futures led to the rejection of the assumption that market event inter-arrival times were independent and identically distributed, challenging the conventional use of independent Poisson processes. Instead, \citet{zhao2010} introduced a Hawkes process, as suggested by \mbox{\citet{bauwens2009}}, to model the arrival rate of market events as a function of recent order arrival rates and quantities. This approach better represented periods of clustered high and low arrival rates, aligning more closely with empirical observations \mbox{\citep{ellul2001determinants, hall2006}}. Similarly, \citet{toke2011} replaced the Poisson processes in the \citet{cont2010stochastic} model with multivariate mutually-exciting Hawkes processes, tailoring them for each type of market event. More recently, \citet{wu2022queue} integrated the queue-reactive model with  Hawkes processes, developing a model that captures both the excitation dynamics and adjusts intensities in accordance with the queue size.


In our study, we follow a methodology akin to that of \citet{zhao2010} and \citet{toke2011}, aiming to compare it to the different versions of the queue-reactive model. To this end, we calibrate a Hawkes process to develop our simulation framework. Our parametrization strategy is as follows:

\begin{itemize}
\item We treat all events occurring at the best limits as a single, six-dimensional multivariate process. This approach accounts for three events at each best price, allowing us to capture not only the excitation between events within the same queue but also the interdependencies between the bid and ask sides of the market.
\item Exponential kernels are employed to model the time-dependent influence of past events on future occurrences. Which means, the memory kernel functions are of the form $h_{ij}(t) =  \alpha_{ij}e^{-\beta_{ij}t}$.
\item We adhere to the same price change paradigm as established in the queue-reactive model, ensuring consistency and comparability in our analyses.
\end{itemize}

For the calibration of this model, we utilized the \textit{tick} package of \citet{2017arXiv170703003B}. Once calibrated, the same package facilitated the simulation of event sequences and interarrival times. These simulations allowed to construct the order flow, which was subsequently incorporated into our matching engine to generate the final simulated order book.

In parallel with the queue-reactive model, we examine two variants of the Hawkes-based model in relation to order sizes. The first variant, referred to as \textbf{Hawkes\_U}, operates on the assumption of unitary order sizes. The second variant, dubbed \textbf{Hawkes\_S}, utilizes the stationary distribution of order sizes. This bifurcation allows for a nuanced examination of the models' performance under different assumptions about market order size models.

\section{Results}
\label{sec:results}
\subsection{Data \& Calibration Paradigm}

The data used for calibration is from Euro-Bund Futures (FGBL)\footnote{Euro-Bund Futures are derivative contracts based on 10-year German government bonds (Bund) and traded on the Eurex Exchange. Each futures contract, with a notional coupon rate of 6\%, has a remaining term to maturity of 8.5 to 10.5 years on the delivery day.} covering active trading days on the year 2021. Data is filtered into periods from 9 AM to 6 PM. The maximum depth considered is five levels on each side. 


\begin{table}[H]
\centering
\begin{tabular}{llllllll}
\cline{1-6}
\multicolumn{1}{l}{\textbf{Level}} & \multicolumn{1}{c}{\textbf{\begin{tabular}[c]{@{}c@{}}\#L\\ $(\times 10 ^6)$\end{tabular}}} & \multicolumn{1}{c}{\textbf{\begin{tabular}[c]{@{}c@{}}\#C\\ $(\times 10 ^6)$\end{tabular}}} & \multicolumn{1}{c}{\textbf{\begin{tabular}[c]{@{}c@{}}\#M\\ $(\times 10 ^4)$\end{tabular}}} & \multicolumn{1}{c}{\textbf{AES}} & \multicolumn{1}{c}{\begin{tabular}[c]{@{}c@{}}\textbf{AIT}\\ (ms)\end{tabular}} & \multicolumn{1}{c}{} & \multicolumn{1}{c}{} \\ \cline{1-6}
1                    & 119.6                                                                             & 115.6                                                                             & 756                                                                              & 5.92                             & 56                                                                              &                      &                      \\ \cline{1-6}
2                    & 24.09                                                                             & 23.84                                                                             & 2.88                                                                             & 5.15                             & 272                                                                             &                      &                      \\ \cline{1-6}
3                    & 15.76                                                                             & 15.29                                                                           & 0                                                                                & 4.87                          & 370                                                                             &                      &                      \\ \cline{1-6}
4                    & 10.72                                                                            & 10.45                                                                           & 0                                                                                & 5.98                             & 468                                                                             &                      &                      \\ \cline{1-6}
5                    & 11.12                                                                           & 10.76                                                                           & 0                                                                                & 4.11                            & 523                                                                             &                      &                      \\ \cline{1-6}
\end{tabular}
\caption{Descriptive statistics about events for each level.}
\label{tab:stat_desc}
\end{table}

Table \ref{tab:stat_desc} provides statistics about the data utilized for model calibration. It displays the number of observations for each event type, the Average Event Size (AES), and the Average Inter-event Time (AIT) for each queue. Notably, the table underscores the high frequency of events impacting queues close to the mid-price, especially the best prices.

Additionally, our dataset includes updates to the limit order book and the daily trade history. However, the models necessitate training with order flow, representing the sequence of events in the limit order book. To obtain this order flow, we carefully constructed it by discerning whether each update indicates liquidity provision (via a limit order) or liquidity consumption (through cancel and market orders). Further details on order flow construction are available in \citet{bodor2024stylized}. Additionally, a matching engine was developed to pair orders produced by each model, subsequently reflecting them in a simulated order book.

This data is further pre-processed following the methodology in \citet{huang2015simulating}. Specifically:

\begin{itemize}
\item Queue sizes are quantized based on the average event size. If the size of queue \(i\) is denoted by \(q_i\) and its average event size is \(AES_i\), then the updated queue size is given by \(q_i \leftarrow \lceil \frac{q_i}{AES_i} \rceil\).
\item The data is divided into periods with a constant reference price. The interarrival time counter for events is reset to 0 for all queues when there is a change in the reference price.

\item The parameter that controls the reference price movements is \( \theta = 0.7\). This probability has been tweaked to reproduce the observed ratio of price continuations to price alternations.

\end{itemize}

For model intensity calibration and subsequent LOB simulations, we:

\begin{itemize}
\item Utilize the calibrated intensities to determine both the time of arrival and the type of new orders.
\item For modeling price changes, we adhere to the aforementioned paradigm. However, we refrain from redrawing the LOB from its stationary distribution to exclusively replicate the endogenous market behavior.
\item Upon a change in the reference price, the size of the queue for newly appearing prices is sampled from their respective stationary distributions.
\end{itemize}

\subsection{Stylized Facts}

In our quest to assess the efficiency of different market simulators, we advocate for their comparison against a set of established stylized facts from the market. These facts span micro to macro-level empirical findings, particularly tailored for the Bund futures market. As detailed in \citet{bodor2024stylized}, we delve into an intricate analysis of stylized facts pertinent to German bond futures, spotlighting Schatz, Bobl, Bund, and Buxl, grounded on exhaustive limit order book data. This exploration discerns both commonalities and unique traits across these futures, shedding light on market dynamics and pointing towards the refinement of market simulation tools. Our primary focus is to benchmark our models against the stylized facts observed in the Bund futures market \citep{bodor2024stylized}, as detailed in the following sections.

\subsubsection{Price dynamics and volatility} 

\begin{figure}[H]
    \centering
    \subfloat[\centering Mid-price]{{\includegraphics[width=0.7\textwidth]{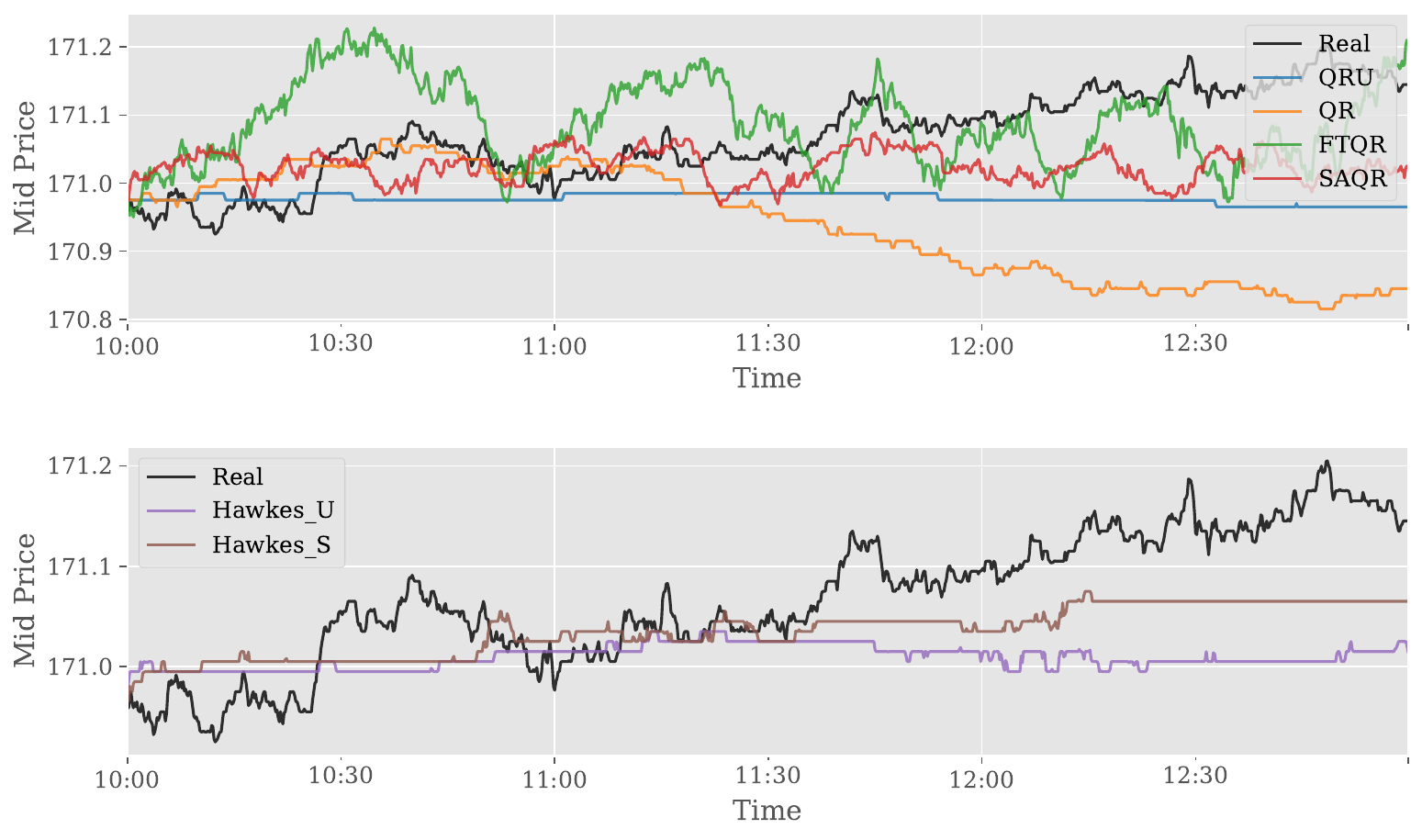}}}%
    \qquad
    \subfloat[\centering Annualized volatility over windows of size 10 minutes]{{\includegraphics[width=0.7\textwidth]{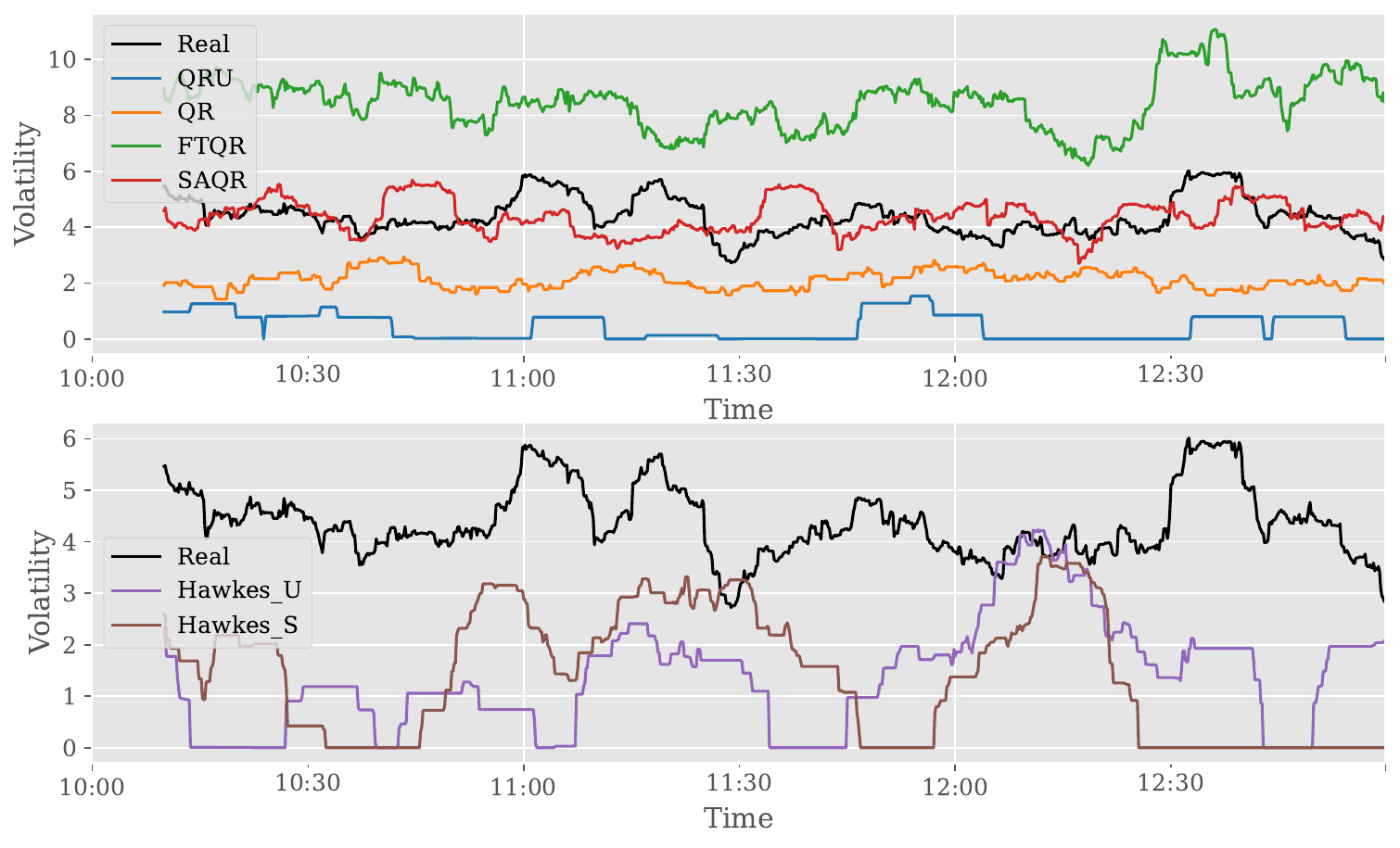} }}%
\caption{Mid-price dynamics and annualized volatility plots (1 second sampling frequency).}%
    \label{fig:price_and_vol}%
\end{figure}

A fundamental feature of an effective simulator is its ability to simulate realistic price dynamics and replicate the volatility inherent to the target market. Figure \ref{fig:price_and_vol} illustrates the price dynamics of both historical data and simulations generated using the different models.


It becomes apparent that the queue-reactive model, particularly the unitary sized variant, exhibits slower dynamics than the actual market, coupled with reduced volatility. Conversely, the FTQR model seems to engender a markedly noisier market environment. Meanwhile, the SAQR model's performance closely aligns with real price patterns. Regarding the Hawkes-based simulation models, they display characteristics similar to those of the standard queue-reactive model. Notably, these simulators also tend to mirror the slower market dynamics observed in the queue-reactive model, contrasting with the quicker pace of real-world markets.



More generally, Table \ref{table:volatility_all} presents a comprehensive comparison of price volatility across different days and distinct periods of the day. This comparison employs two metrics:

\begin{itemize}
    \item Relative difference: \( \text{mean}\left(100\% \times \frac{\sigma_s - \sigma_r}{\sigma_r}\right) \)
    \item Quadratic error: \( \text{mean}\left((\sigma_s - \sigma_r)^2\right) \)
\end{itemize}

Here, \( \sigma_s \) and \( \sigma_r \) denote the annualized volatility of the simulated and real markets, respectively.

The subsequent table delineates the comparative volatility of simulated markets vis-à-vis the real market. Notably, the SAQR model persistently outperforms other models across all periods of the day, reflecting the smallest average relative difference and quadratic error. This sharply contrasts the QRU model, which exhibits considerably diminished volatility relative to the real market.

The performance trajectories of the QR and FTQR models warrant special attention, as they fluctuate based on the time of day. The QR model exhibits closer adherence to real market dynamics during the 10 AM to 2 PM bracket, generally perceived as the ``calm'' periods of the day. Conversely, the FTQR model manifests superior performance from 3 PM to 6 PM, periods known for high market activity driven by events like closing auctions and options expiry.

\begin{table}[H]
\centering
\small
\begin{tabular}{l|cc|cc|cc|}
\cline{2-7}
\multicolumn{1}{c|}{}               & \multicolumn{2}{c|}{9 AM - 6 PM}                                                                                                   & \multicolumn{2}{c|}{10 AM - 2 PM}                                                                                                  & \multicolumn{2}{c|}{3 PM - 6 PM}                                                                                                  \\ \cline{2-7} 
\multicolumn{1}{c|}{}               & \multicolumn{1}{c}{\begin{tabular}[c]{@{}c@{}}Relative \\ Difference\\ (\%)\end{tabular}} & \multicolumn{1}{c|}{\begin{tabular}[c]{@{}c@{}}Quadratic \\ Error\\ \end{tabular}} & \multicolumn{1}{c}{\begin{tabular}[c]{@{}c@{}}Relative \\ Difference\\ (\%)\end{tabular}} & \multicolumn{1}{c|}{\begin{tabular}[c]{@{}c@{}}Quadratic \\ Error\\ \end{tabular}} & \multicolumn{1}{c}{\begin{tabular}[c]{@{}c@{}}Relative \\ Difference\\ (\%)\end{tabular}} & \multicolumn{1}{c|}{\begin{tabular}[c]{@{}c@{}}Quadratic \\ Error\\ \end{tabular}} \\ \hline
\multicolumn{1}{|l|}{\textbf{QRU}}  & -91                                                            & 18.37                                                      & -90                                                            & 14.97                                                      & -91                                                            & 21.50                                                      \\ \hline
\multicolumn{1}{|l|}{\textbf{QR}}   & -52                                                            & 7.34                                                       & -48                                                           & 5.17                                                       & -54                                                            & 9.56                                                       \\ \hline
\multicolumn{1}{|l|}{\textbf{FTQR}} & 69                                                             & 8.88                                                       & 82                                                             & 10.71                                                      & 49                                                             & 8.12                                                       \\ \hline
\multicolumn{1}{|l|}{\textbf{SAQR}} & \textbf{1}                                                              & \textbf{1.35}                                                       & \textbf{8}                                                           & \textbf{0.82}                                                       & \textbf{-3}                                                       & \textbf{2.33}                                                       \\ \hline
\multicolumn{1}{|l|}{\textbf{Hawkes\_U}} & -72                                                         & 12.46                                                      & -70                                                           & 9.64                                                       & -73                                                            & 15.16                                                      \\ \hline
\multicolumn{1}{|l|}{\textbf{Hawkes\_S}} & -69                                                         & 11.70                                                      & -67                                                            & 8.96                                                       & -71                                                            & 14.33                                                      \\ \hline
\end{tabular}
\caption{Statistics about volatility of simulated markets compared with real market on different periods of the day.}
\label{table:volatility_all}
\end{table}




\subsubsection{Transactions magnitude and market activity}

\begin{figure}[H]
	\centering
	\includegraphics[width=0.7\textwidth]{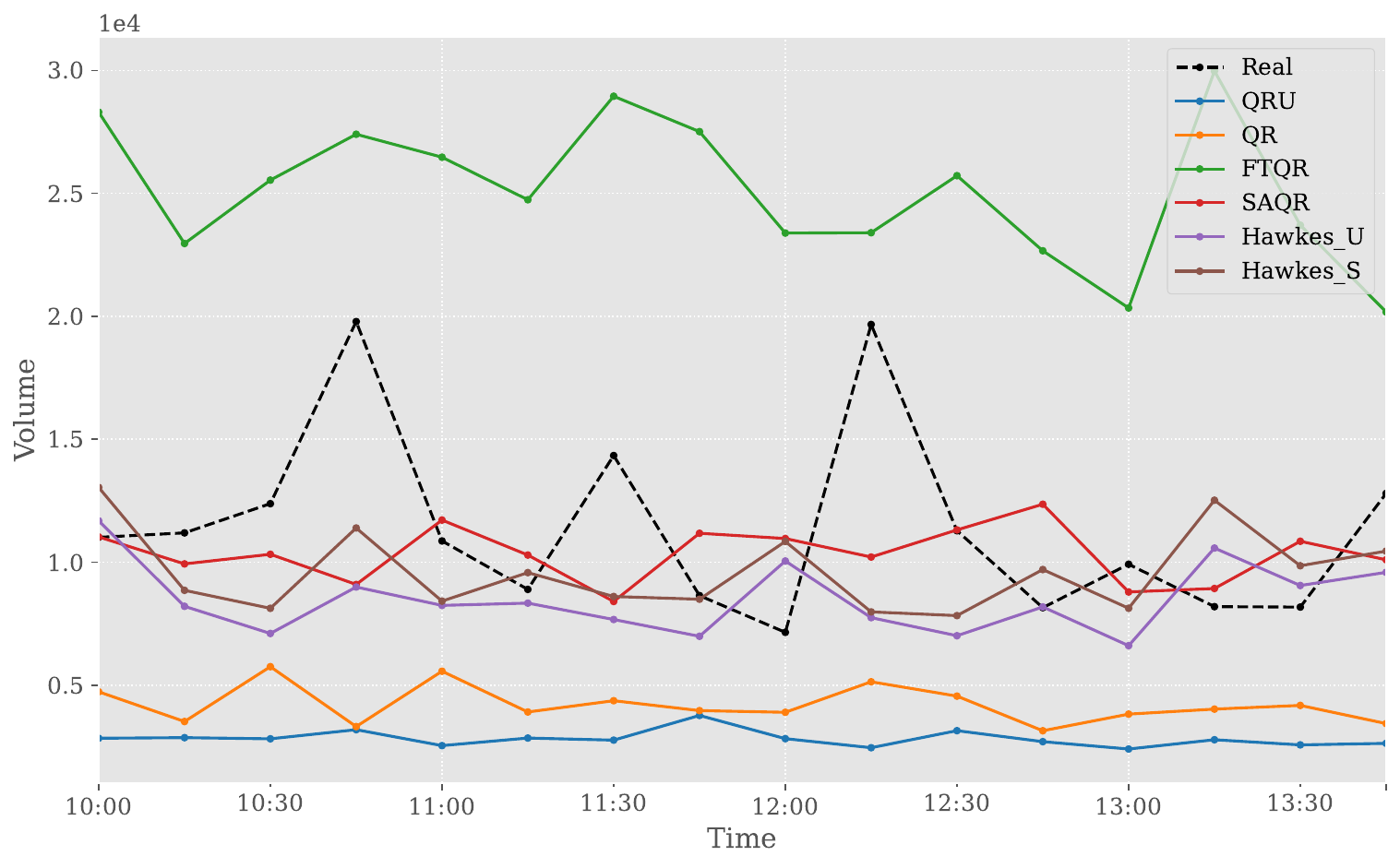}
	\caption{Traded volumes on windows of size 10 minutes.}
	\label{fig:Traded_volumes_on_windows_of_size_10_minutes}
\end{figure}

An essential aspect of limit order book simulators is their ability to accurately capture the magnitude of transactions and orders. This entails the capability of the simulator to replicate a similar order of magnitude of traded volumes within a given time window. As illustrated in Figure \ref{fig:Traded_volumes_on_windows_of_size_10_minutes}, which shows that the SAQR model exhibits the highest proficiency in replicating traded volumes during typical trading periods of the day. Both the QR and QRU models demonstrate a relatively low transactions volume. However, the FTQR model exhibits a significantly higher rate. The Hawkes-based models are also quite accurate at reproducing a relatively similar market activity, showcasing the importance of excitation between events for the verification of this stylized fact.

\begin{table}[H]

\center
\small
\begin{tabular}{l|cc|cc|cc|}
\cline{2-7}
\multicolumn{1}{c|}{}               & \multicolumn{2}{c|}{9 AM - 6 PM}                                                                                                                                                             & \multicolumn{2}{c|}{10 AM - 2 PM}                                                                                                                                                            & \multicolumn{2}{c|}{3 PM - 6 PM}                                                                                                                                                           \\ \cline{2-7} 
\multicolumn{1}{c|}{}               & \multicolumn{1}{c}{\begin{tabular}[c]{@{}c@{}}Relative \\ Difference\\ (\%)\end{tabular}} & \multicolumn{1}{c|}{\begin{tabular}[c]{@{}c@{}}Quadratic \\ Error\\ ($\times 10^9$)\end{tabular}} & \multicolumn{1}{c}{\begin{tabular}[c]{@{}c@{}}Relative \\ Difference\\ (\%)\end{tabular}} & \multicolumn{1}{c|}{\begin{tabular}[c]{@{}c@{}}Quadratic \\ Error\\ ($\times 10^9$)\end{tabular}} & \multicolumn{1}{c}{\begin{tabular}[c]{@{}c@{}}Relative \\ Difference\\ (\%)\end{tabular}} & \multicolumn{1}{c|}{\begin{tabular}[c]{@{}c@{}}Quadratic \\ Error\\ ($\times 10^9$)\end{tabular}} \\ \hline
\multicolumn{1}{|l|}{\textbf{QRU}}  & -80                                                                                        & 2.02                                                                                      & -76                                                                                        & 1.32                                                                                      & -82                                                                                        & 2.81                                                                                      \\ \hline
\multicolumn{1}{|l|}{\textbf{QR}}   & -70                                                                                        & 1.55                                                                                      & -65                                                                                        & 0.95                                                                                      & -74                                                                                        & 2.26                                                                                      \\ \hline
\multicolumn{1}{|l|}{\textbf{FTQR}} & 49                                                                                       & 0.85                                                                                      & 76                                                                                       & 1.38                                                                                      & \textbf{30}                                                                                       & \textbf{0.63}                                                                                      \\ \hline
\multicolumn{1}{|l|}{\textbf{SAQR}} & \textbf{-26}                                                                               & \textbf{0.36}                                                                             & \textbf{-13}                                                                               & \textbf{0.14}                                                                             & -36                                                                               & \textbf{0.62}                                                                             \\ \hline
\multicolumn{1}{|l|}{\textbf{Hawkes\_U}} & -40                                                                                        & 0.60                                                                                      & -28                                                                                        & 0.33                                                                                      & -48                                                                                       & 0.97                                                                                      \\ \hline
\multicolumn{1}{|l|}{\textbf{Hawkes\_S}} & -32                                                                                        & 0.41                                                                                      & -19                                                                                        & 0.21                                                                                      & -41                                                                                        & 0.80                                                                                      \\ \hline

\end{tabular}
\caption{Statistics about traded volumes of simulated during different periods of the day.}
\label{table:trades_all}
\end{table}

Table \ref{table:trades_all} provides a comprehensive comparison between the traded volumes of the simulated market and the real market. Consistently, the SAQR model emerges as the closest to real data, exhibiting the smallest quadratic error across all day periods, followed by the Hawkes-based models, that show also quite good results, with a notable improvement of the Hawkes\_S model over Hawkes\_U. Interestingly, the QR model outperforms the FTQR model during the periods from 10 AM to 2 PM, while the FTQR model appears to be better suited for replicating the high market activity observed from 3 PM to 6 PM.

\subsubsection{Order book queue size distribution}

An additional noteworthy empirical observation pertains to the distribution of available liquidity in the market. Numerous studies have not only acknowledged the consistency of this distribution but also reported the successful fit of the Gamma distribution to it \citep{abergel2016limit, vyetrenko2020get, bodor2024stylized}.

\begin{figure}[H]
	\centering
	\includegraphics[width=1\textwidth]{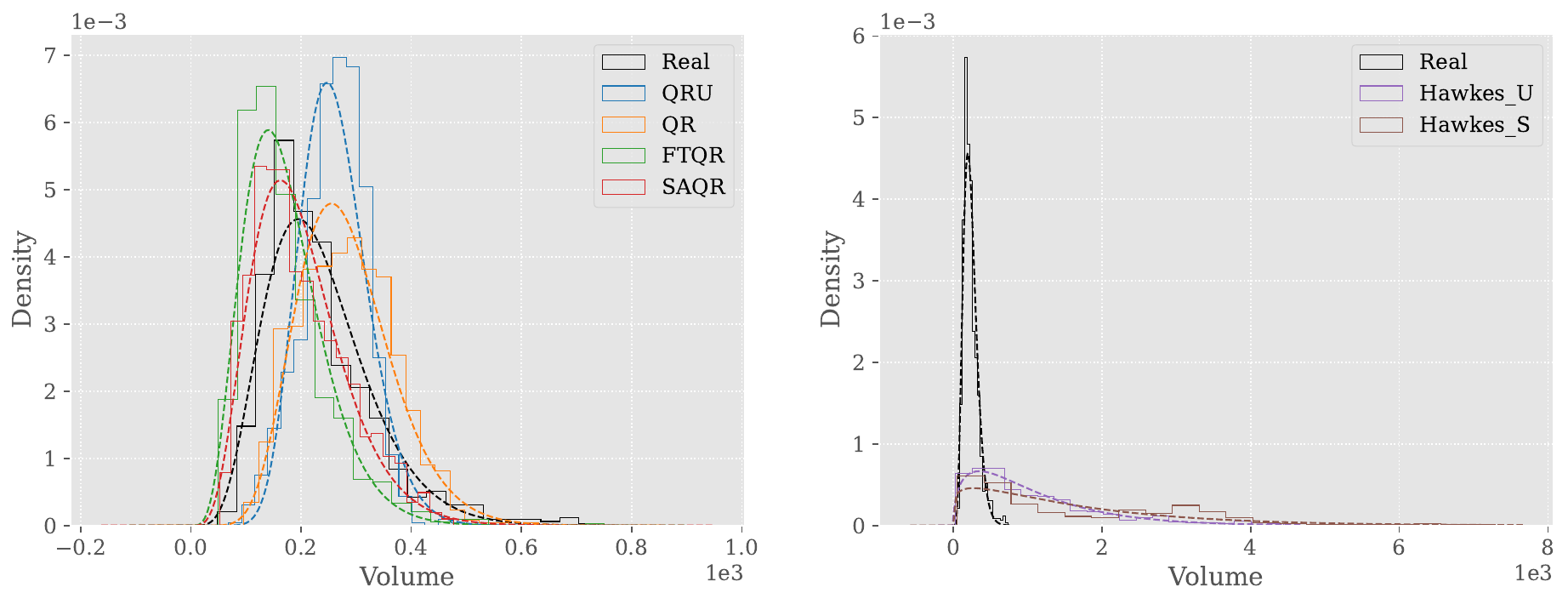}
	\caption{Distribution of order book volumes and Gamma law fit.}
	\label{fig:Distribution_of_order_book_volumes_and_Gamma_law_fit}
\end{figure}

In Figure \ref{fig:Distribution_of_order_book_volumes_and_Gamma_law_fit}, the empirical distribution of queue sizes at the most favorable levels is effectively captured. All models fit accurately the Gamma distribution, highlighting their capability in modeling queue dynamics. Notably, the SAQR and FTQR models demonstrate a particularly strong alignment with actual market data, underscoring their accuracy in reflecting real market conditions.
In contrast, the Hawkes-based models do not mimic the queue size distribution as accurately. They tend to simulate many limit orders, regardless of the queue's actual size. This results in an unrealistic pile-up of orders, which does not match what we usually see in real markets. These observations highlight the strengths of the queue-reactive model. It adjusts the flow of orders – both incoming and outgoing – based on the size of the queue. This empirical validation of the queue-reactive model's ergodicity, which was theoretically established for unitary sizes in \citet{huang2015simulating}  and extended to a general order size model in \citet{lehalle2021optimal}, underscores its efficacy in replicating actual market dynamics, especially in terms of queue size distribution. 

In order to compare these distributions across multiple days of data, we adopt the Kolmogorov-Smirnov statistic as a distance between distributions. The Kolmogorov-Smirnov (KS) test is a nonparametric test used to determine the similarity between two probability distributions. The statistic of the KS test is defined as:
\begin{equation*}
D_n = \sup_x |F_n(x) - F(x)|
\end{equation*}
where $F_n(x)$ is the empirical cumulative distribution function (CDF) of the observed data and $F(x)$ is the CDF of the reference distribution or theoretical model. The value of $D_n$ lies between 0 and 1. The smaller the value of $D_n$, the more similar the observed data distribution is to the reference distribution. Therefore, the KS statistic can be used to measure the similarity between two distributions.

Table \ref{tab:gamma_volumes_dist} presents the descriptive statistics of the Kolmogorov-Smirnov statistic, comparing the real and simulated queue size distributions from different days of calibration data. The table indicates that both the SAQR and QR models perform well on average, despite the fact that the volume distributions generated by these two models are quite distinct. This discrepancy can be attributed to the variability in market behavior, which aligns more closely with the QR distribution on some days and the SAQR distribution on others. Regardless, both models can be considered to validate the stylized fact. In contrast, Hawkes-based models exhibit very wide distribution distances from the real distribution, further confirming that these models suffer from the queue size divergence problem, where, intrinsically, there is no guarantee of a return to equilibrium, unlike the queue reactive model.

\begin{table}[H]
\centering
\small
\begin{tabular}{l|c|c|c|c|c|c|}
\cline{2-7}
                           & \textbf{QRU} & \textbf{QR} & \textbf{FTQR} & \textbf{SAQR} & \textbf{Hawkes\_U} & \textbf{Hawkes\_S} \\ \hline
\multicolumn{1}{|l|}{mean} & 0.32         & 0.22        & 0.28          & \textbf{0.20} & 0.63               & 0.69               \\ \hline
\multicolumn{1}{|l|}{std}  & 0.09         & 0.11        & 0.08          & 0.06          & 0.05               & 0.04               \\ \hline
\multicolumn{1}{|l|}{min}  & 0.23         & 0.14        & 0.05          & 0.08          & 0.57               & 0.64               \\ \hline
\multicolumn{1}{|l|}{25\%} & 0.27         & 0.15        & 0.26          & 0.15          & 0.60               & 0.66               \\ \hline
\multicolumn{1}{|l|}{50\%} & 0.30         & 0.18        & 0.30          & 0.21          & 0.62               & 0.68               \\ \hline
\multicolumn{1}{|l|}{75\%} & 0.31         & 0.22        & 0.33          & 0.26          & 0.65               & 0.71               \\ \hline
\multicolumn{1}{|l|}{max}  & 0.68         & 0.61        & 0.37          & 0.28          & 0.78               & 0.81               \\ \hline
\end{tabular}
\caption{Kolmogorov-Smirnov statistic of the distribution of queue sizes in simulated markets compared to the real market, description over multiple days of calibration data.}
\label{tab:gamma_volumes_dist}
\end{table}

\subsubsection{Signature plot}
\begin{figure}[H]
	\centering
	\includegraphics[width=0.7\textwidth]{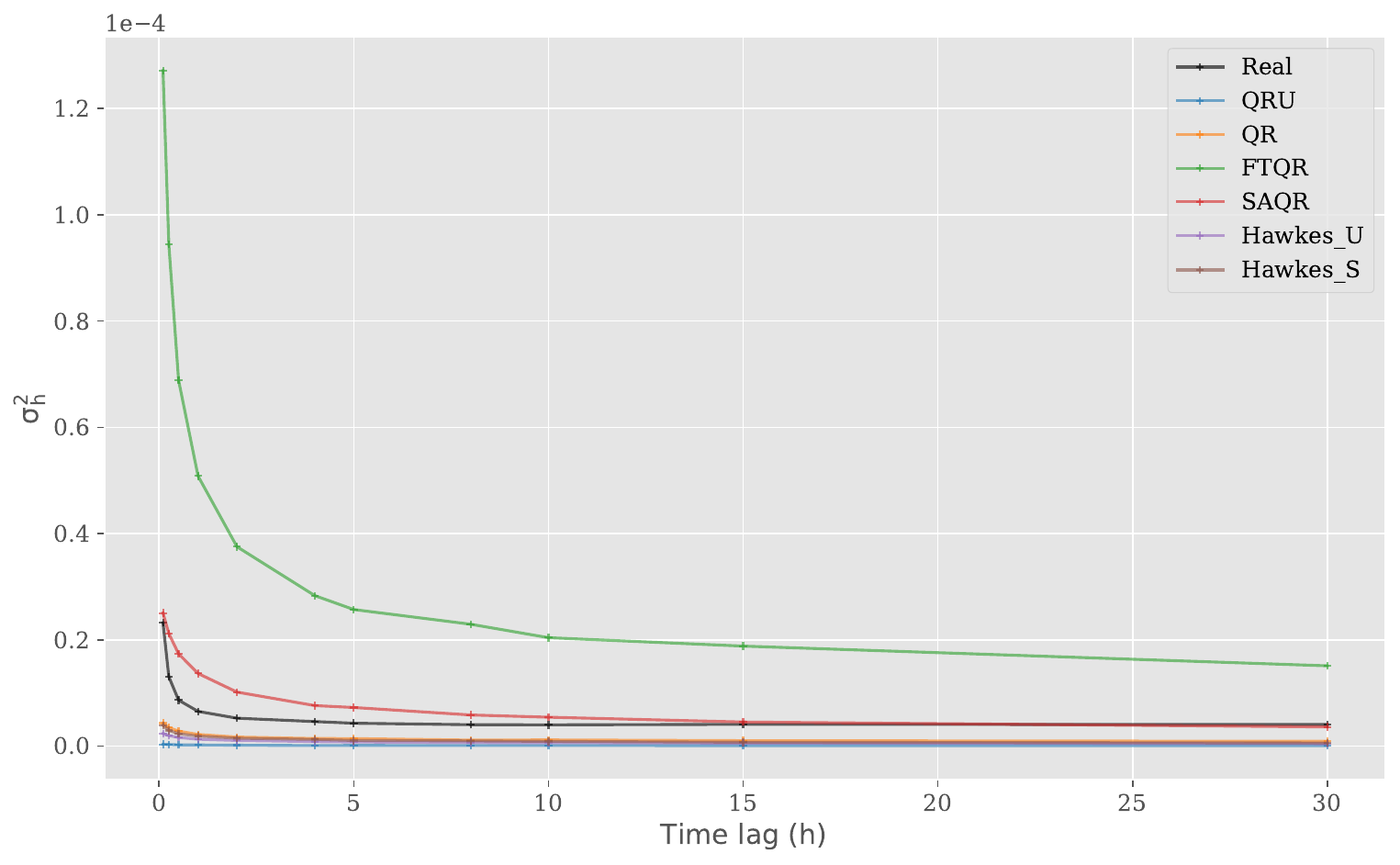}
	\caption{Signature plot.}
	\label{fig:Signature_plot}
\end{figure}

The signature plot of a price is defined as the variance of price increments at a given lag value $h$, divided by $h$. In other words, it is defined as $ \sigma^2_{h} = \frac{\mathbb{V}\left(P(t+h) - P(t)\right)}{h}$, $P(t)$ being the mid-price of the asset at time $t$. It is interesting to observe the evolution of this metric with the value of $h$, as this highlights the transition of price movements at different time scales. Figure \ref{fig:Signature_plot} shows this transition for an example of one real day and simulated markets. While all the markets seem to exhibit the same shape of decrease in volatility as the time lag value increases, the SAQR model is the best at capturing the real signature plot at different time scales. 

More generally, the following table provides more statistics regarding the comparison of the signature plot of real data from different days to the signature plot of simulated data, the distance between curves is defined as the sum of difference of values for each lag $h$. The table shows that the QRU, QR, Hawkes\_U and Hawkes\_S models generally underestimate the volatility while the FTQR heavily overestimates it. The SAQR is the model with the least quadratic error and least bias, confirming its ability to capture the real volatility at different scales. However, it is important to note that while the SAQR model excels in terms of quadratic error and bias, it still struggles to perfectly capture the exact decaying profile of the real data, especially at smaller time scales.

\begin{table}[H]
\centering
\small
\begin{tabular}{l|l|l|}
\cline{2-3}
& \multicolumn{1}{c|}{\textbf{\begin{tabular}[c]{@{}c@{}}Relative \\ Difference  \\ (\%)\end{tabular}}} & \multicolumn{1}{c|}{\textbf{\begin{tabular}[c]{@{}c@{}} $\mathbf{L^2}$ \\ Norm \\ $\times 10^{-5}$\end{tabular}}} \\ \hline
\multicolumn{1}{|l|}{\textbf{QRU}}  & -9.80                                                                             & 2.89                                                                              \\ \hline
\multicolumn{1}{|l|}{\textbf{QR}}   & -7.55                                                                             & 2.28                                                                              \\ \hline
\multicolumn{1}{|l|}{\textbf{FTQR}} & 41.32                                                                             & 9.99                                                                              \\ \hline
\multicolumn{1}{|l|}{\textbf{SAQR}} & \textbf{4.43}                                                                     & \textbf{1.05}                                                                     \\ \hline
\multicolumn{1}{|l|}{\textbf{Hawkes\_U}} & -8.57 & 2.09 \\ \hline
\multicolumn{1}{|l|}{\textbf{Hawkes\_S}} & -7.91 & 1.95 \\ \hline
\end{tabular}
\caption{Statistics over multiple days of calibration data - signature plot values.}
\end{table}

\subsubsection{Distribution of returns}

\begin{figure}[H]
    \centering
\subfloat[\centering All models]{{	\includegraphics[width=0.45\textwidth]{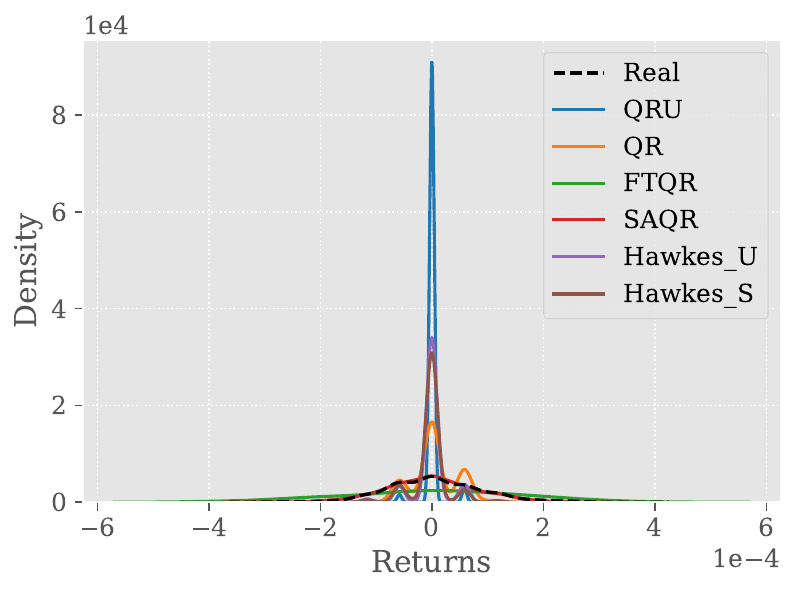} }}%
    \qquad
      \subfloat[\centering Closest models only]{{	\includegraphics[width=0.45\textwidth]{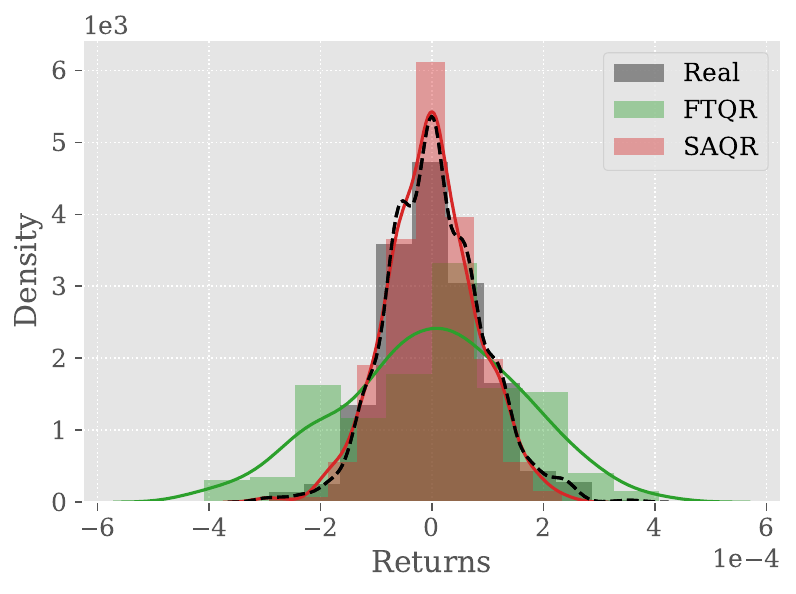}}}%
\caption{Distribution of logarithmic returns $\log(\frac{P(t+\tau)}{P(t)})$ of real and simulated data.}
	\label{fig:Histogram_of_returns}
\end{figure}

Figure \ref{fig:Histogram_of_returns} displays a comparison of daily return distributions across various simulated markets and the real market. Part (a) of the figure compares the distributions of all models against the real market, while part (b) focuses on the best-performing models for enhanced clarity. This comparison reveals that the SAQR model's distribution of returns closely resembles that of the real market. In part (a), it is noticeable that the QR, QRU, Hawkes\_U, and Hawkes\_S models generate returns that are too concentrated around zero, reflecting a slower volatility than what is seen in the actual market. On the other hand, part (b) illustrates how the SAQR model's return distribution closely mirrors the real market. This finding is corroborated by Table \ref{table:returns_ks}, which indicates that the SAQR and FTQR models have return distributions most similar to those of the real market, with the SAQR model being the closest.

\begin{table}[H]
\centering
\small
\begin{tabular}{l|c|c|c|c|c|c|}
\cline{2-7}
& \textbf{QRU} & \textbf{QR} & \textbf{FTQR} & \textbf{SAQR} & \textbf{Hawkes\_U} & \textbf{Hawkes\_S} \\ \hline
\multicolumn{1}{|l|}{mean} & 0.52 & 0.35 & 0.21 & \textbf{0.16} & 0.43 & 0.42 \\ \hline
\multicolumn{1}{|l|}{std} & 0.04 & 0.05 & 0.05 & 0.03 & 0.03 & 0.03 \\ \hline
\multicolumn{1}{|l|}{min} & 0.43 & 0.25 & 0.12 & 0.11 & 0.39 & 0.37 \\ \hline
\multicolumn{1}{|l|}{25\%} & 0.49 & 0.31 & 0.17 & 0.14 & 0.41 & 0.40 \\ \hline
\multicolumn{1}{|l|}{50\%} & 0.51 & 0.35 & 0.20 & 0.15 & 0.43 & 0.42 \\ \hline
\multicolumn{1}{|l|}{75\%} & 0.55 & 0.37 & 0.23 & 0.17 & 0.46 & 0.44 \\ \hline
\multicolumn{1}{|l|}{max} & 0.60 & 0.44 & 0.35 & 0.25 & 0.50 & 0.48 \\  \hline
\end{tabular}
\caption{Kolmogorov-Smirnov statistic of the distribution of returns in simulated markets compared to the real market. Description over multiple days of calibration data}
\label{table:returns_ks}
\end{table}




\subsubsection{Order sizes distribution}

\begin{figure}[H]
	\centering
	\includegraphics[width=0.7\textwidth]{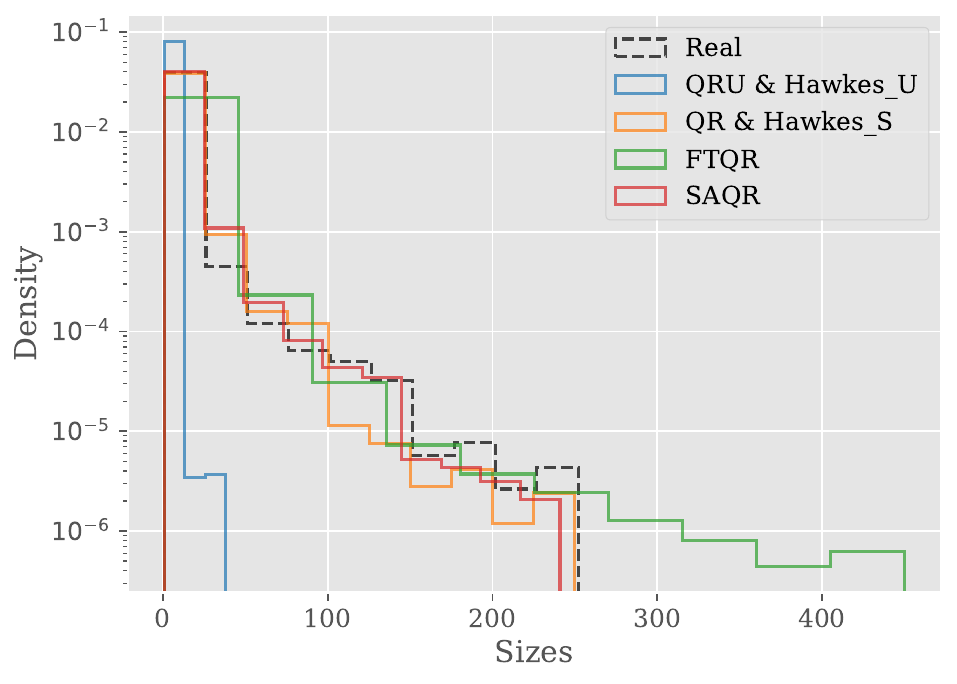}
	\caption{Distribution of order sizes (the y-axis is in logarithmic scale).}
	\label{fig:Histogram_of_order_sizes}
\end{figure}

Figure \ref{fig:Histogram_of_order_sizes} provides insights into the distribution of order sizes in both real and simulated markets. Several observations can be made:

\begin{itemize}

\item As anticipated, the QRU and Hawkes\_U models perform poorly in reproducing this stylized fact. All orders in these models are unitary and clustered around the average event size (AES), resulting in a limited range of order sizes, and completely different distribution.

\item The QR and Hawkes\_S models exhibit a good fit to the historical distribution of order sizes. This outcome is expected since these models sample, by construction, order sizes from the stationary distribution of real market order sizes.

\item Conversely, the SAQR model demonstrates a strong capability in reproducing the distribution of order sizes. It is important to emphasize that, in contrast to the QR model, this capability is not inherent to the SAQR model's design. The order sizes in the SAQR model are drawn based on a distribution conditional on the queue size. This feature allows the SAQR model to capture the empirical order size distribution more accurately, accounting for the realism of the queue sizes dynamics of this model.

\item The FTQR model, while similar to the QR model in terms of order size picking paradigm, presents a distribution slightly shifted towards larger sizes. This shift is primarily due to the introduction of the two event types \emph{cancel\_all} and \emph{market\_all} which encourage the placement of larger orders. This could also justify the increased volatility and activity observed in this model.

\end{itemize}




\subsubsection{Order book shape}
\begin{figure}[H]
	\centering
	\includegraphics[width=0.7\textwidth]{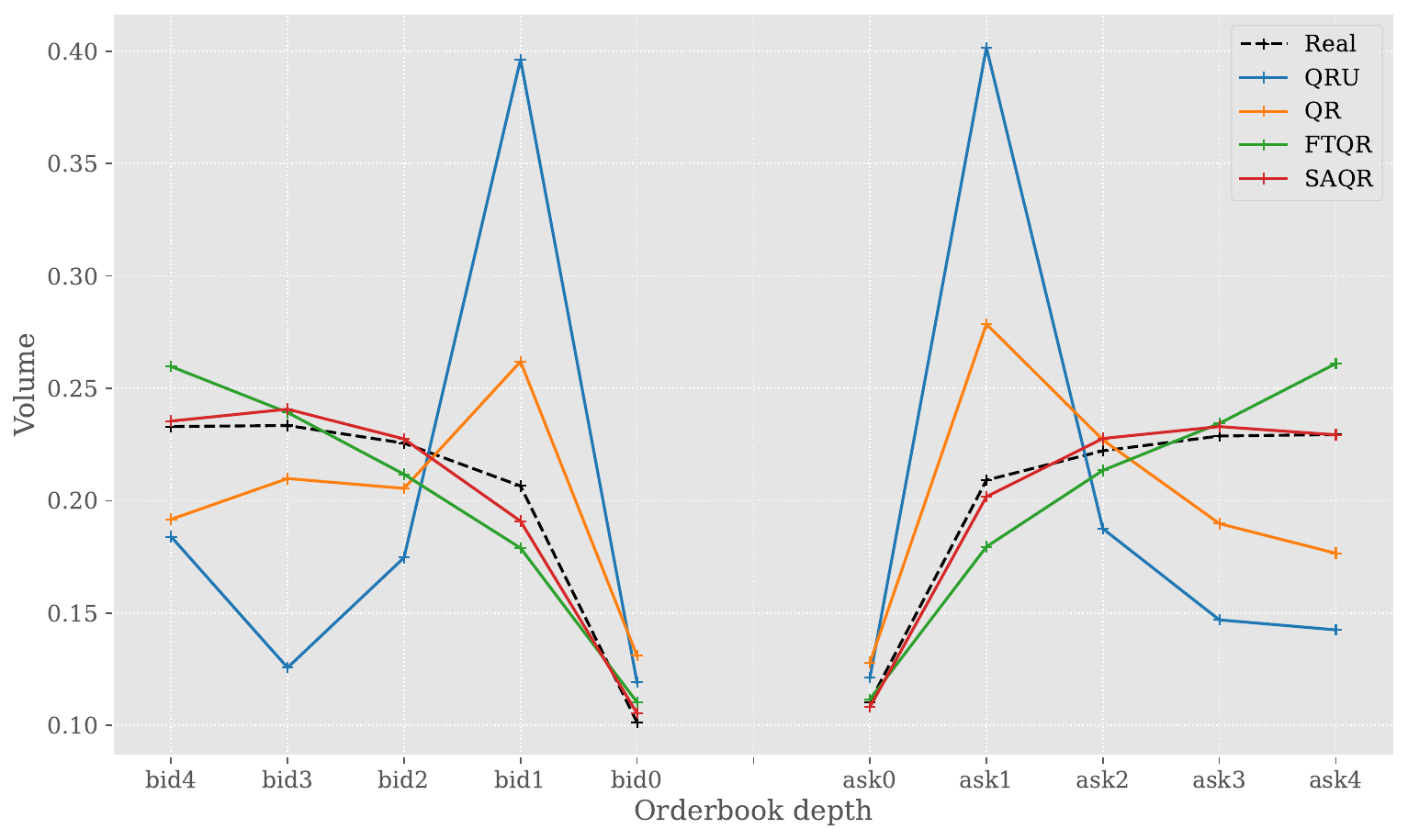}
	\caption{Order book mean volumes (averaged).}
	\label{fig:Orderbook_Mean_volumes_(averaged)}
\end{figure}

Figure \ref{fig:Orderbook_Mean_volumes_(averaged)} illustrates the average volume available at each queue up to the fifth level, normalized by the average of all queue sizes. This figure indicates that the limit order book is generally more populated at deeper levels, a phenomenon also observed on other markets \citep{abergel2016limit, chakraborti2011econophysics}, especially on the Bund (see \cite{bodor2024stylized}).

This stylized fact is particularly evident in the FTQR and SAQR models. However, it appears less applicable to the QRU and QR models. This difference underscores the superior ability of the FTQR and SAQR models to accurately simulate queue size distributions at various price depths. The Hawkes-based models are absent in the figure as they simulate only best prices order flow.

\subsubsection{Summary}
Additional stylized facts are detailed in \ref{appendix:ComplementarySF}. Tables \ref{tab:SFs_names} and \ref{tab:models_perfs} present a comprehensive list of stylized facts examined in our study, along with a performance summary for each model concerning these facts. We have used a color-coded system for clarity: green signifies that a model successfully captures a stylized fact, orange denotes partial accuracy, red indicates a miss, and black dash means irrelevant. The SAQR model emerges as a clear frontrunner, corroborating the most stylized facts and thereby underscoring its proficiency in mimicking real market dynamics. Conversely, the QRU and Hawkes\_U models lag behind, emphasizing the significance of incorporating order sizes in market simulations. To sum up, accurately modeling order sizes is instrumental in enhancing the authenticity of simulated markets and bolstering the efficacy of the queue-reactive model.

\begin{table}[H]
	\centering
	\begin{tabular}{|c|l|}
	\hline
	\rowcolor[HTML]{00C057} 
	\textbf{Index} & \multicolumn{1}{c|}{\cellcolor[HTML]{00C057}\textbf{Stylized fact name}} \\ \hline
	1            & Distribution of order sizes                                      \\ \hline
	2            & Power-law of order sizes distribution                                 \\ \hline
	3            & Signature plot                                             \\ \hline
	4            & Distribution of available volumes in the queue           \\ \hline
	5            & Price dynamics and volatility                                       \\ \hline
	6            & Long range dependency                                      \\ \hline
	7            & Distribution of returns                                  \\ \hline
	8            & Order book shape                                            \\ \hline
	9           & Absence of autocorrelation                               \\ \hline
	10           & Traded volumes in a fixed window                         \\ \hline
	11           & Weibull fit of interarrival time of trades              \\ \hline
12           & Excitation between events              \\ \hline
	\end{tabular}
	\caption{Table of stylized facts.}
  \label{tab:SFs_names}
\end{table}


\begin{table}[H]
\centering
\begin{tabular}{|l|c|c|c|c|c|c|c|c|c|c|c|c|}
\hline
\rowcolor[HTML]{00C057} 
\multicolumn{1}{|l|}{\cellcolor[HTML]{00C057}\textit{\textbf{Model}}} & \textit{\textbf{1}}                & \textit{\textbf{2}}                & \textit{\textbf{3}}                & \textit{\textbf{4}}                & \textit{\textbf{5}}                & \textit{\textbf{6}}                & \textit{\textbf{7}}                & \textit{\textbf{8}}                & \textit{\textbf{9}}               & \textit{\textbf{10}}               & \textit{\textbf{11}} & \textit{\textbf{12}}             \\ \hline
\cellcolor[HTML]{00C057}\textit{\textbf{QRU}}                         & {\color[HTML]{FF0000} \textbf{✗}}  & {\color[HTML]{FF0000} \textbf{✗}}  & {\color[HTML]{FF0000} \textbf{✗}}  & {\color[HTML]{BF8F00} \textbf{☑}}  & {\color[HTML]{FF0000} \textbf{✗}}  & {\color[HTML]{FF0000} \textbf{✗}}  & {\color[HTML]{FF0000} \textbf{✗}}  & {\color[HTML]{FF0000} \textbf{✗}}  & {\color[HTML]{00B050} \textbf{☑}} & {\color[HTML]{FF0000} \textbf{✗}}  & {\color[HTML]{00B050} \textbf{☑}} & {\color[HTML]{FF0000} \textbf{✗}} \\ \hline
\cellcolor[HTML]{00C057}\textit{\textbf{QR}}                          & {\color[HTML]{00B050} \textbf{☑}} & {\color[HTML]{00B050} \textbf{☑}} & {\color[HTML]{FF0000} \textbf{✗}}  & {\color[HTML]{00B050} \textbf{☑}} & {\color[HTML]{FF0000} \textbf{✗}}  & {\color[HTML]{00B050} \textbf{☑}} & {\color[HTML]{FF0000} \textbf{✗}}  & {\color[HTML]{FF0000} \textbf{✗}}  & {\color[HTML]{00B050} \textbf{☑}} & {\color[HTML]{FF0000} \textbf{✗}}  & {\color[HTML]{00B050} \textbf{☑}} & {\color[HTML]{FF0000} \textbf{✗}} \\ \hline
\cellcolor[HTML]{00C057}\textit{\textbf{FTQR}}                        & {\color[HTML]{FF0000} \textbf{✗}}  & {\color[HTML]{00B050} \textbf{☑}} & {\color[HTML]{FF0000} \textbf{✗}}  & {\color[HTML]{BF8F00} \textbf{☑}} & {\color[HTML]{FF0000} \textbf{✗}}  & {\color[HTML]{00B050} \textbf{☑}} & {\color[HTML]{FF0000} \textbf{✗}}  & {\color[HTML]{00B050} \textbf{☑}} & {\color[HTML]{00B050} \textbf{☑}} & {\color[HTML]{FF0000} \textbf{✗}}  & {\color[HTML]{BF8F00} \textbf{☑}} & {\color[HTML]{FF0000} \textbf{✗}} \\ \hline
\cellcolor[HTML]{00C057}\textit{\textbf{SAQR}}                        & {\color[HTML]{00B050} \textbf{☑}} & {\color[HTML]{00B050} \textbf{☑}} & {\color[HTML]{BF8F00} \textbf{☑}} & {\color[HTML]{00B050} \textbf{☑}} & {\color[HTML]{00B050} \textbf{☑}} & {\color[HTML]{00B050} \textbf{☑}} & {\color[HTML]{00B050} \textbf{☑}} & {\color[HTML]{00B050} \textbf{☑}} & {\color[HTML]{00B050} \textbf{☑}} & {\color[HTML]{00B050} \textbf{☑}} & {\color[HTML]{BF8F00} \textbf{☑}} & {\color[HTML]{FF0000} \textbf{✗}} \\ \hline
\cellcolor[HTML]{00C057}\textit{\textbf{Hawkes\_U}}                  & {\color[HTML]{FF0000} \textbf{✗}}  & {\color[HTML]{FF0000} \textbf{✗}} & {\color[HTML]{FF0000} \textbf{✗}}  & {\color[HTML]{FF0000} \textbf{✗}}  & {\color[HTML]{FF0000} \textbf{✗}} & {\color[HTML]{BF8F00} \textbf{☑}} & {\color[HTML]{FF0000} \textbf{✗}} & - & {\color[HTML]{00B050} \textbf{☑}} & {\color[HTML]{00B050} \textbf{☑}} & {\color[HTML]{BF8F00} \textbf{☑}} & {\color[HTML]{00B050} \textbf{☑}} \\ \hline
\cellcolor[HTML]{00C057}\textit{\textbf{Hawkes\_S}}                  & {\color[HTML]{00B050} \textbf{☑}} & {\color[HTML]{00B050} \textbf{☑}} & {\color[HTML]{FF0000} \textbf{✗}}  & {\color[HTML]{FF0000} \textbf{✗}} & {\color[HTML]{FF0000} \textbf{✗}} & {\color[HTML]{00B050} \textbf{☑}} & {\color[HTML]{FF0000} \textbf{✗}} & - & {\color[HTML]{00B050} \textbf{☑}} & {\color[HTML]{00B050} \textbf{☑}} & {\color[HTML]{BF8F00} \textbf{☑}} & {\color[HTML]{00B050} \textbf{☑}} \\ \hline
\end{tabular}
\caption{Summary of stylized facts verified by each model.}
\label{tab:models_perfs}
\end{table}

\section{Conclusion and Perspectives}

In this paper, we explore adaptations of the queue-reactive model to accurately simulate the dynamics of limit order books, using empirical data, with a focus on German bond futures. Our research uniquely incorporates and demonstrates the critical importance of order sizes, alongside order types and arrival rates, in realistically modeling limit order book flows. This methodological enhancement is validated by aligning closely with numerous market stylized facts, emphasizing the model's ability to accurately replicate real market behaviors. Moving forward, we aim to further refine the model by integrating additional dynamics such as event excitation and intra-day seasonality, among others, ensuring a balance between simplicity and comprehensive market representation.

\section*{Acknowledgements}

We extend our sincere thanks to the first author's PhD supervisor, Pr. Olivier Guéant, for his invaluable guidance and insights throughout this project. Our gratitude is also due to the Vietnam Symposium in Banking and Finance 2023 for the opportunity to present and discuss our work with esteemed researchers. Finally, we would like to express our appreciation to Pr. Charles-Albert Lehalle and Pr. Mathieu Rosenbaum for the constructive discussions about their paper and model, and for the precious advice on future research directions.

\section*{Disclosure statement}
No potential conflict of interest was reported by the authors.

\section*{Declaration of Funding}
This work was conducted as part of a CIFRE thesis in collaboration with the Sorbonne Economics Centre, with financial support from BNP Paribas Arbitrage. The contribution of BNP Paribas Arbitrage has been pivotal to the research, ensuring access to essential data, providing necessary funding, and supplying resources and materials to conduct the research work. Despite these contributions, BNP Paribas Arbitrage had no influence on the study's design, analysis, interpretation of the results, or the decision-making process regarding this publication.

 \bibliographystyle{elsarticle-num-names}
 \bibliography{cas-refs}
 
\newpage

\appendix


\section{Complementary Stylized Facts Results}
\label{appendix:ComplementarySF}

\subsection{Autocorrelation of price returns}
\begin{figure}[H]
\centering
\includegraphics[width=0.8\textwidth]{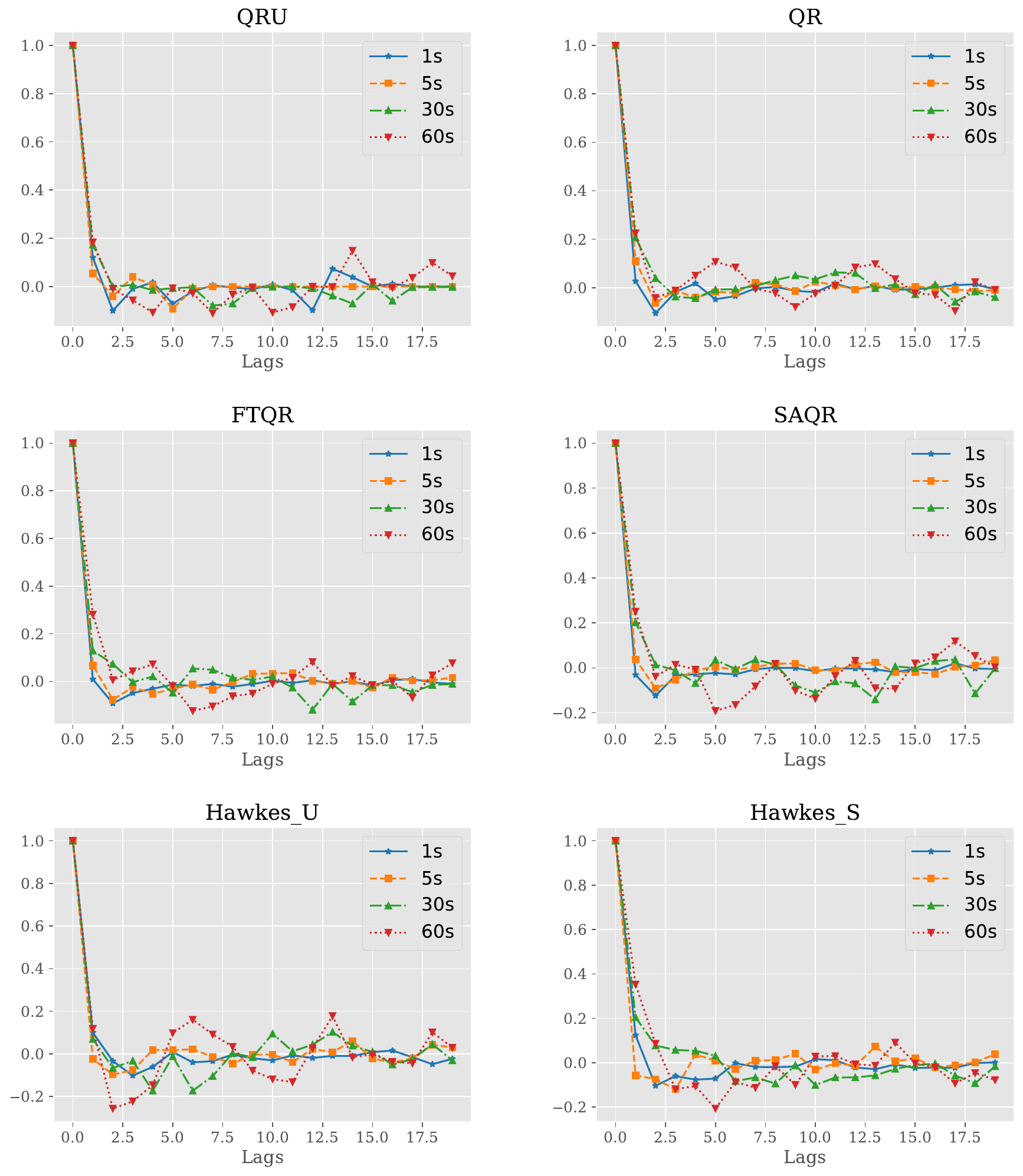}
\caption{Autocorrelation of returns for different sampling frequencies.}
\label{fig:absence_of_correlation}
\end{figure}

Figure \ref{fig:absence_of_correlation} displays the autocorrelation of returns for different sampling frequencies and for the six models, it shows that there is generally no autocorrelation between returns at different time scales.

\newpage
\subsection{Long Range Dependence}
\begin{figure}[H]
\centering
\includegraphics[width=0.8\textwidth]{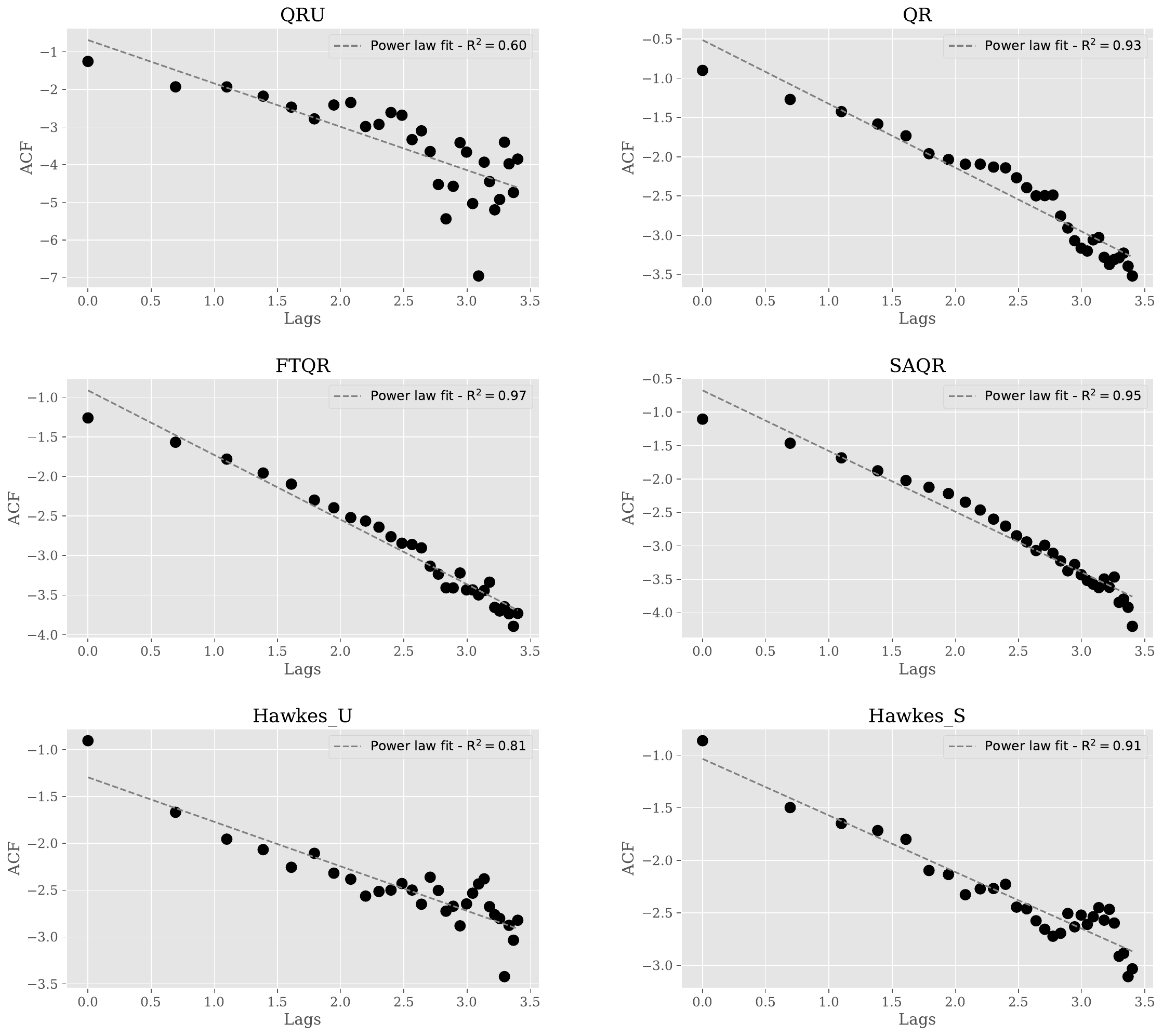}
\caption{Long range dependence.}
\label{fig:Long_range_dependence}
\end{figure}

While a realistic market should show no correlation between returns, it is commonly observed in different markets that the correlation of absolute returns shows a slow decay \citep{cont2001empirical}. Moreover, this decay follows a power-law distribution. Figure \ref{fig:Long_range_dependence} shows how simulated markets exhibit this slow decay of absolute returns correlation, with a pretty good fit of power-law distribution, except for the QRU and Hawkes\_U models.

\newpage
\subsection{Weibull Fit of Interarrival Time of Trades}
\begin{figure}[H]
\centering
\includegraphics[width=0.75\textwidth]{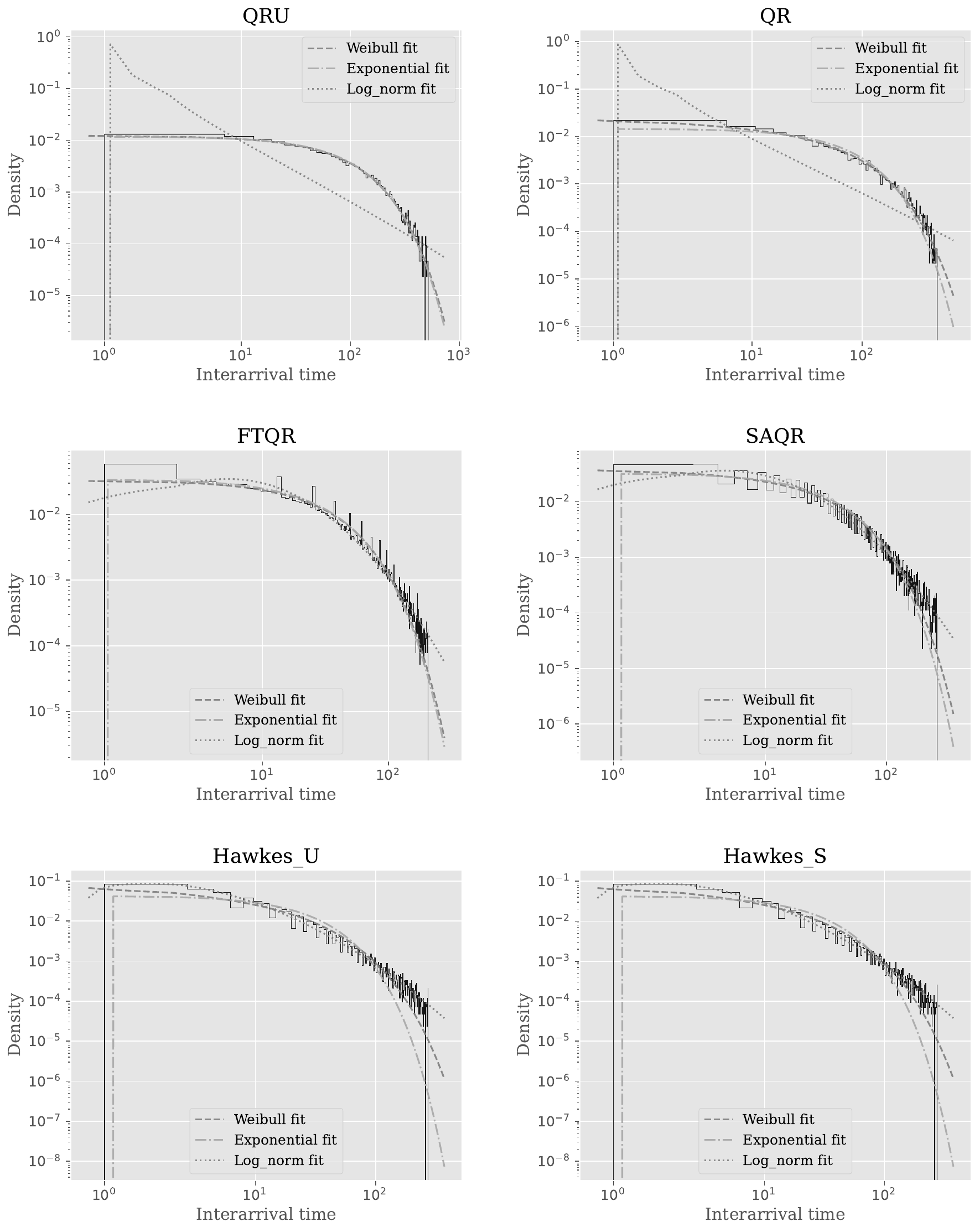}
\caption{Distribution of interarrival times of market orders.}
\label{fig:Distribution_of_interarrival_times_of_trade_orders}
\end{figure}

Another known stylized fact is the Weibull fit of interarrival times of market orders. Results in Figure \ref{fig:Distribution_of_interarrival_times_of_trade_orders} show that all the models verify this stylized fact, as the Weibull distribution is the closest fit. Moreover, this result aligns with previous empirical observations of the Bund \citep{bodor2024stylized}, and is similar to findings for other assets such as equity \citep{abergel2016limit}. It is worth noting that this stylized fact is an important consideration in the analysis of financial markets.

\newpage
\subsection{Excitation between events}
\begin{figure}[H]
\centering
\includegraphics[width=0.7\textwidth]{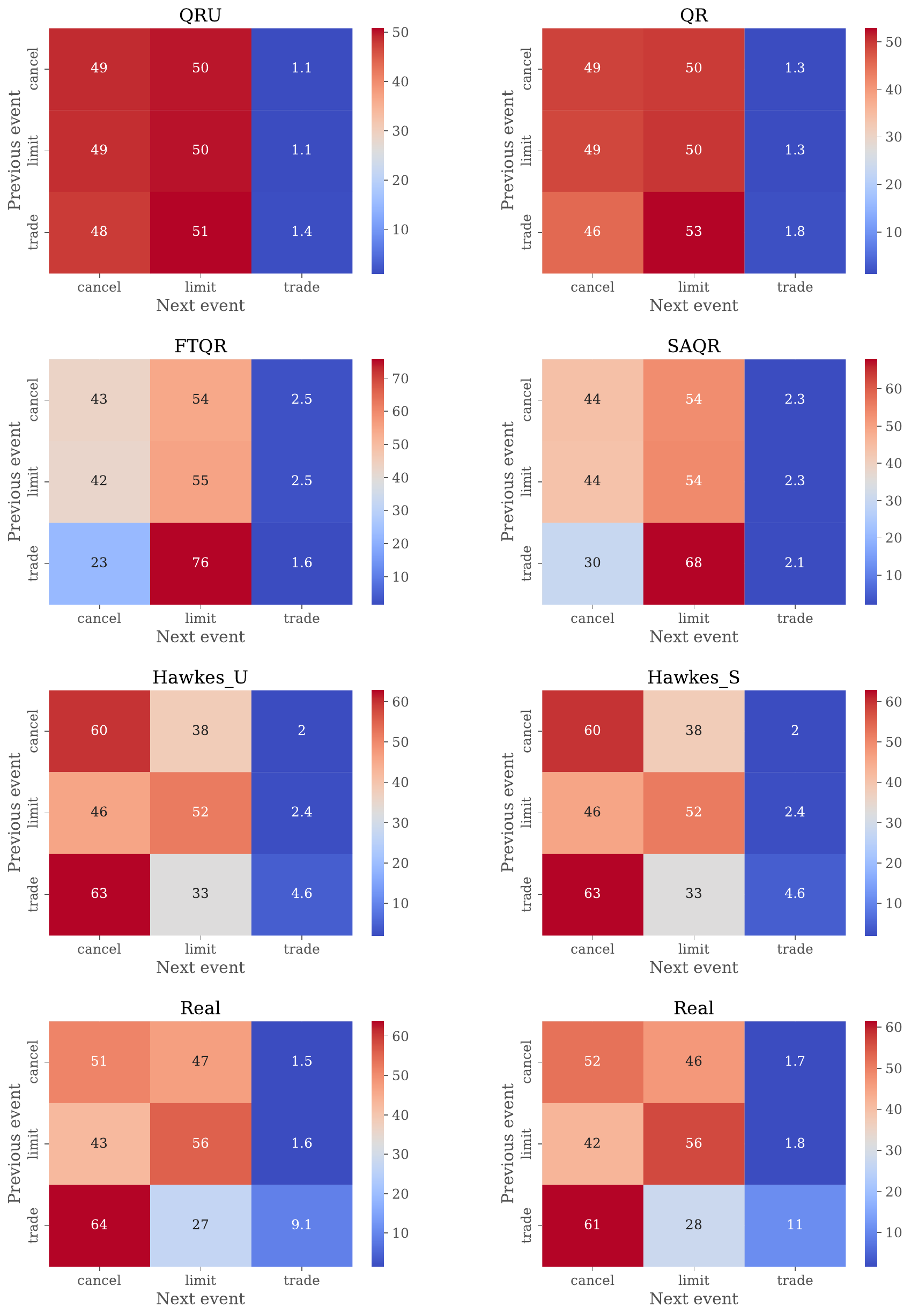}
\caption{Transition probabilities between events.}
\label{fig:excitation_matrices_probabilites}
\end{figure}

Figure \ref{fig:excitation_matrices_probabilites} presents the matrix of conditional probabilities of an event $E_{t^+}$ (columns) given a past event $E_{t^-}$ (rows), denoted as $\probP \left( E_{t^+} | E_{t^-} \right)$. The matrices reveal that the queue-reactive models inadequately replicate the empirical matrices, as evidenced by nearly uniform rows in the matrix. This indicates that the occurrence of a past event does not significantly influence the likelihood of future events in these models. In contrast, Hawkes-based models demonstrate a notable accuracy in reproducing the excitation between events, attributable to their design and the inclusion of a memory component.

\newpage
\subsection{Power-law of Order Sizes}
\begin{figure}[H]
\centering
\includegraphics[width=0.8\textwidth]{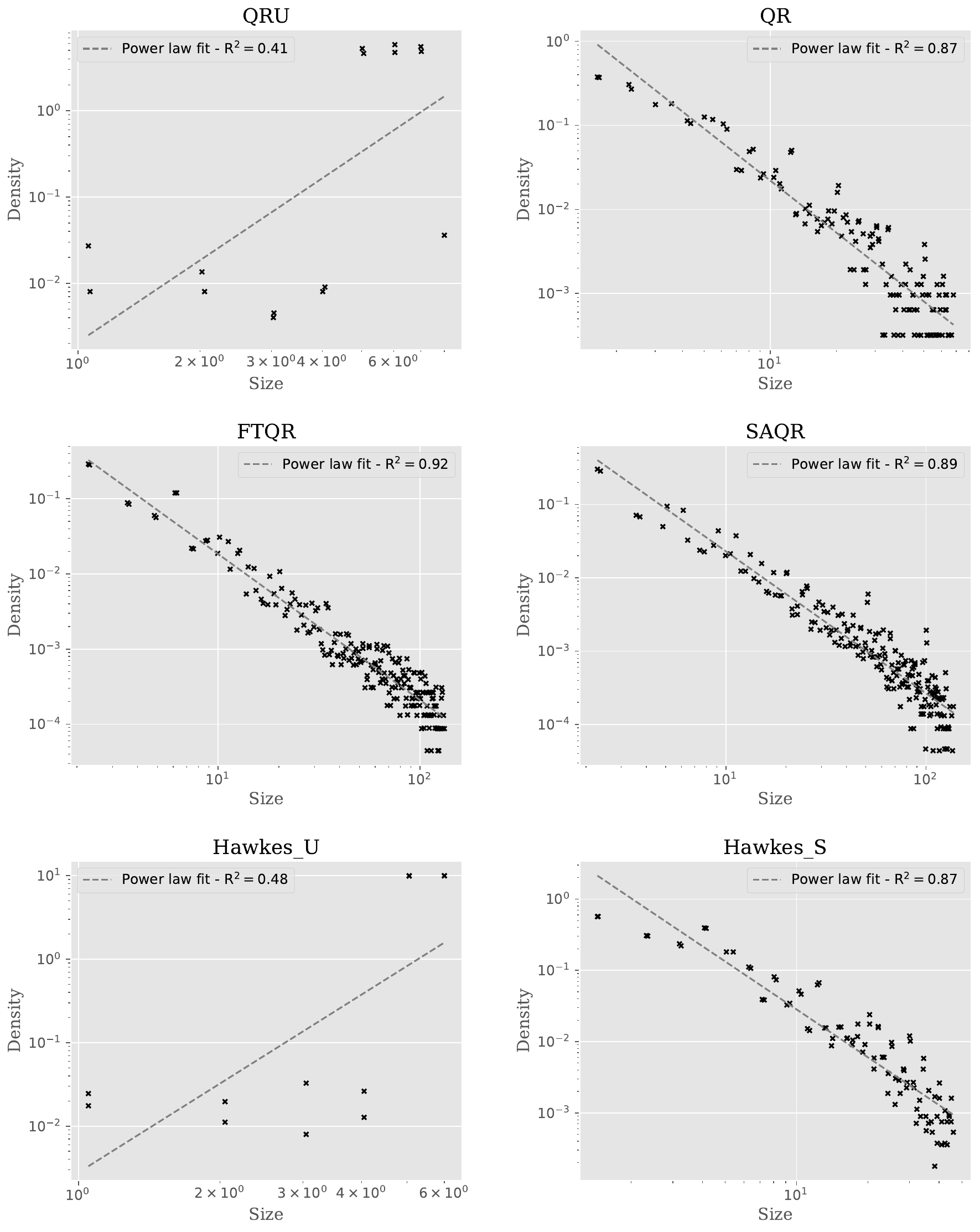}
\caption{Distribution of trade order sizes and power-law fit.}
\label{fig:Distribution_of_trade_order_sizes_and_power_law_fit}
\end{figure}

The power-law fit of order sizes is also one of the stylized facts that is verified in multiple markets \citep{chakraborti2011econophysics, abergel2016limit, bouchaud2018trades}. In particular, Bund futures market exhibits this behavior \citep{bodor2024stylized}. Figure \ref{fig:Distribution_of_trade_order_sizes_and_power_law_fit} shows that all the models satisfy this stylized fact, except for the QRU and Hawkes\_U models, mainly because of the unitary nature of the order sizes.

\end{document}